\documentclass[a4paper,12pt,preprint,amsmath,showpacs,nofootinbib,groupedaddress,tightenlines]{revtex4-1}
\usepackage{color}
\usepackage{slashed}
\usepackage{graphicx}   % need for figures
\usepackage{slashed}
\usepackage[lofdepth,lotdepth]{subfig}

\newcommand{\dd}{\mathrm{d}}

\newcommand{\als}{\alpha_s}
\newcommand{\ep}{\epsilon}

\newcommand{\nn}{\nonumber}

\newcommand{\ws}{\widetilde{s}}

\newcommand{\orda}{\mathcal{O}(\alpha_s)}
\newcommand{\ordb}{\mathcal{O}(\alpha^2_s)}

\def\mb#1{{\boldsymbol #1}}
\def\ord#1{{\mathcal{O}(#1)}}

\begin{document}

\preprint{SLAC-PUB-16050}
\preprint{SMU-HEP-14-06}

\title{Electroweak prodution of top-quark pairs in $e^+e^-$ annihilation at NNLO in QCD: the vector contributions}

\author{Jun Gao}
\affiliation{Department of Physics, Southern Methodist University, Dallas, TX 75275-0181, USA}
\author{Hua Xing Zhu}
\affiliation{SLAC National Accelerator Laboratory, Stanford University, Stanford, CA 94309, USA}

\begin{abstract}
\noindent
We report on a calculation of the vector current contributions to the electroweak
production of top quark pairs in $e^+e^-$ annihilation at
next-to-next-to-leading order in Quantum Chromodynamics. Our setup is fully
differential and can be used to calculate any infrared-safe
observable. The real emission contributions are handled by a next-to-next-to-leading order
generalization of the phase-space slicing method. We demonstrate the power of our technique by considering its application to various inclusive and exclusive observables.
\end{abstract}

\maketitle
%\allowdisplaybreaks
%\allowdisplaybreaks
\section{I\lowercase{ntroduction}}
\label{sec:intro}

Continuum electroweak production of top quark pairs at future linear colliders is of considerable interest because it allows for a precise
measurement of the top quark forward-backward asymmetry.
This observable is of particular importance because it is expected to severely constrain anomalous couplings which could potentially appear in the top quark sector~\cite{hep-ex/0509008}. 
In the near future, due to the extremely clean environment expected at proposed $e^+ e^-$ colliders, it should be possible to measure the top quark forward-backward asymmetry to a precision of approximately $1\%$~\cite{1005.1756}.

At an $e^+e^-$ collider, top quark pairs are primarily produced via the electroweak process
\begin{align}
  e^+e^- \to \gamma^* / Z^* \to t \bar{t} \,.
\end{align}
In this paper, we shall only concern ourselves with the next-to-next-to-leading order~(NNLO) radiative corrections to the above process in Quantum Chromodynamics (QCD) mediated by an off-shell photon ($\gamma^*$).
In other words, we treat the vector current contributions to the production of a top-antitop pair. Complete results including the axial-vector contributions ({\it i.e.} that due to off-shell $Z$ boson exchange) will be presented elsewhere.

The calculation of QCD radiative corrections to heavy-quark pair production in $e^+e^-$ annihilation has a long history. 
Full next-to-leading order~(NLO) QCD corrections were first computed in ref.~\cite{PHRVA.D25.1218} and, a short time later, NLO electroweak effects were considered in ref.~\cite{NUPHA.B365.24}. 
NLO QCD corrections to top quark pair production including the subsequent top quark decays were presented in ref.~\cite{Schmidt:1995mr} and
NLO QCD corrections to top quark spin correlations were computed in refs.~\cite{hep-ph/9807209} and~\cite{hep-ph/9901205}.
Total cross sections are known to NNLO in the threshold expansions~\cite{hep-ph/9712222,hep-ph/9712302,hep-ph/9801397,hep-ph/0001286,hep-ph/9508274} and high-energy expansions~\cite{Gorishnii:1986pz,PHLTA.B248.359,hep-ph/9406299,hep-ph/9710413,hep-ph/9704222}. 
Results for the forward-backward asymmetry are also known in the small mass approximation~\cite{NUPHA.B391.3,hep-ph/9809411,hep-ph/9905424}. In the near future, the threshold cross section at NNNLO will also be available~\cite{Beneke:2013jia,Marquard:2014pea,Beneke:2014qea}.
Somewhat surprisingly, although a great deal of theoretical progress has been made over the years, exact NNLO QCD calculations for fully differential $e^+e^- \to t\bar{t}$ observables remain a challenge and are still missing from the literature.

A fully differential NNLO QCD calculation is naturally split up into three distinct parts, depending on the number of particles that appear in the final state relative to leading order:
a) purely virtual two-loop or squared one-loop corrections, 
b) one-loop, single-emission real-virtual corrections, 
and c) double-emission double-real corrections. 
For $e^+e^- \to t\bar{t}$, significant progress has been made in recent years towards the calculation of each of these three pieces. 
NLO QCD corrections to heavy quark pair production in association with one additional jet were computed in refs.~\cite{hep-ph/9703358,hep-ph/9705295,hep-ph/9708350,hep-ph/9709360,hep-ph/9905276}. 
The two-loop heavy quark form factor was first obtained in refs.~\cite{hep-ph/0406046,hep-ph/0412259,hep-ph/0504190} and then confirmed some time later by an independent calculation~\cite{0905.1137}.
In fact, for quite some time, the only outstanding problem was to construct an efficient framework for the combination of the ingredients described above into an infrared-safe Monte Carlo event generator. 

For generic processes, this is highly non-trivial due to the fact that, in phase space regions where soft and/or collinear
limits are approached, the real-virtual and double-real contributions develop soft and/or collinear divergences which must be extracted before a Monte Carlo integration over phase space can be carried out.
At NLO, this is relatively straightforward to do and both phase-space slicing~\cite{Fabricius:1981sx,Kramer:1986mc,Baer:1989jg,PHRVA.D46.1980,hep-ph/9302225,hep-ph/0102128,Keller:1998tf} and subtraction~\cite{Ellis:1980wv,Mangano:1991jk,Kunszt:1992tn,hep-ph/9512328,hep-ph/9605323,hep-ph/0201036}
techniques which solve the problem were worked out long time ago. However, as is clear from the massive amount of literature on the subject~\cite{hep-ph/0004013,hep-ph/0302180,hep-ph/0306248,hep-ph/0311311,
hep-ph/0402265,hep-ph/0402280,hep-ph/0403057,hep-ph/0409088,hep-ph/0411399,hep-ph/0502226,hep-ph/0505111,hep-ph/0603182,hep-ph/0609042,hep-ph/0609043,hep-ph/0703012,0710.0346,0802.0813,0807.0509,
0807.3241,PHLTA.B666.336,0807.0514,0903.2120,0904.1145,0905.4390,0912.0374,1003.2824,JHEPA.1002.089,1003.4451,1005.0274,1011.1909,1011.6631,1101.0642,1105.0530,1107.4037,1107.1164,1110.2368,1110.2375,
1112.4736,1204.5201,1207.5779,1207.6546,1210.2808,1210.5059,1301.4693,1301.7133,1301.7310,1302.6216,1303.6254,1309.6887,1309.7000,1310.3993,1401.7754,1404.6493,1404.7116,1405.2219,Anastasiou:2014nha},
analogous techniques at NNLO are considerably more complicated to develop and complete solutions took much longer to emerge.  For example,
in the important case of massless dijet production, it took more than a decade for the first physical predictions to appear~\cite{1301.7310,1310.3993} from the time that the relevant two-loop virtual amplitudes
were first calculated~\cite{hep-ph/0001001,hep-ph/9905323,hep-ph/9909506,hep-ph/0003261,hep-ph/0010212,hep-ph/0011094,hep-ph/0012007,hep-ph/0101304,hep-ph/0102201,hep-ph/0104178}. As a result of significant theoretical efforts during the past decade, a number of important ``benchmark processes'' are now known to NNLO~\cite{1204.5201,1301.7310,1302.6216,1303.6254,1310.3993,1405.2219}.

The goal of this paper is to study fully differential NNLO QCD corrections to $e^+e^- \to t\bar{t}$ using a higher-order generalization of the phase-space slicing method.
While we constrain ourselves in this paper to present results for the vector current contributions by themselves,
the formalism developed here can, if desired, readily be used to calculate the contributions coming from the exchange of an off-shell $Z$ boson.
This paper is organized as follows. In Section~\ref{sec:slice}, we describe our calculational method in detail. 
In Section~\ref{sec:num}, we present numerical results for various inclusive and differential observables and, whenever possible, compare them to the existing literature. 
Finally, we conclude in Section~\ref{sec:conc}.

\section{P\lowercase{hase-space slicing at} NNLO}
\label{sec:slice}

We explain in detail our generalization of phase-space slicing
method in dealing with the specific process $e^+e^- \to t\bar{t}$ at
NNLO. As mentioned before, there are three distinct parts contribute to the cross section at $\ord{\als^2}$,
\begin{align}
  \sigma^{(2)} =& \int\! \dd \Phi_{t\bar{t},0} \sum_{\mathrm{spin,color}}\left[ 2\Re\left( \mathcal{M}^{(0)}_{e^+e^-\to t\bar{t}} \left( \mathcal{M}^{(2)}_{e^+e^-\to t\bar{t}}\right)^* \right) +
\left| \mathcal{M}^{(1)}_{e^+e^-\to t\bar{t}} \right|^2 \right]
\nn
\\
&
+ \int\! \dd \Phi_{t\bar{t},1} \sum_{\mathrm{spin,color}} \left[ 2 \Re
  \left(\mathcal{M}^{(0)}_{e^+e^-\to t\bar{t} g}
    \left(\mathcal{M}^{(1)}_{e^+e^-\to t\bar{t} g} \right)^*\right)
\right]  
\nn
\\
&
+ \int\! \dd \Phi_{t\bar{t},2} \sum_{\mathrm{spin,color}}\left[ \left|\mathcal{M}^{(0)}_{e^+e^- \to t
    \bar{t} gg}\right|^2 + \left|\mathcal{M}^{(0)}_{e^+e^- \to t
    \bar{t} q\bar{q}}\right|^2  \right],
\label{eq:sigma2}
\end{align}
where 
\begin{align}
  \dd \Phi_{t\bar{t},n} =& \frac{1}{2s \times 2^2} \left(
    \frac{\dd^{D-1} p_t}{2E_t (2\pi)^{D-1}}\right) \left(
    \frac{\dd^{D-1} p_{\bar t}}{2E_{\bar t} (2\pi)^{D-1}}\right)
\nn
\\
&
\times \prod^n_{i=1} \left(
    \frac{\dd^{D-1} p_{i}}{2E_{i} (2\pi)^{D-1}}\right)(2\pi)^D
  \delta^{(D)}\left( Q - p_t - p_{\bar t} - \sum_{i=1}^n p_i \right)
\end{align}
is the phase space volume element in $D=4-2\ep$ dimension, divided by the flux factor and
initial state spin average factor. Here $s=Q^2 = (p_{e^+} +
p_{e^-})^2$ is the center-of-mass energy
square. $\mathcal{M}^{(i)}_{e^+e^- \to t\bar{t}\dots}$ denotes the
$i$-loop amplitude for $e^+e^-\to t\bar{t}$ plus zero, one, or two
additional massless partons. Note that when $\sqrt{s}>4m_t$, the
channel for the production of $t\bar{t} t\bar{t}$ is open. However,
these additional contributions are themselves infrared finite due to
the mass of top quark, and can
be dealt with separately. In the following discussion, we will neglect
these contributions. Also for the vector contributions, we only consider diagrams with top quarks coupling directly to photon. The diagrams with photon coupling to a bottom or light quarks and the top quark produced via gluon splitting are numerically small~\cite{Hoang:1994it,hep-ph/9605311}.
 Though the bottom triangle diagrams are needed and must be included to cancel the axial anomaly in the axial vector case~\cite{hep-ph/9710413,hep-ph/0504190}.

The first, second, and third terms on the RHS of Eq.(\ref{eq:sigma2})
represent respectively the double-virtual, real-virtual, and double-real contributions. The double-virtual contributions contain explicit
quadratic poles in $\ep$, originating from loop corrections when the
gluons are soft. Thanks to
Bloch-Nordsieck and Kinoshita-Lee-Nauenberg theorem, the infrared
divergences will be cancelled by those in the
real-virtual and double-real contributions. However, such
cancellation is non-trivial because the
infrared divergences in the real-virutal and double-real contributions
can only be made explicit \emph{after} phase space integral. It is
therefore necessary to perform the phase space integral in $D$
dimension to regulate potential infrared divergences. This fact makes
the calculation of real-virtual and double-real contributions
difficult. 

The singular region in the phase space is relatively simple for the
real-virtual corrections, where the matrix elements are singular only
when the energy of the final state gluon approaches zero. For the
double-real contributions the singular region is much more
involved. First, the matrix elements are singular in the double
un-resolved region, where the energies of both the final state partons
approach zero. Second, the matrix elements are also singular even in
the single un-resolved region, where only one of the final state gluon is
soft, or the final state massless partons become
collinear. Fortunately, the singularities due to single un-resolved
region is well understood, as they are the same one encounters in NLO
QCD calculation. We therefore only need to deal with the double
un-resolved region. To isolate the phase space singularities in this
region, we introduce a phase-space slicing parameter $\tau$, which is
proportional to the total energy of QCD radiations in the final state,
$\tau=2(\sqrt{s}-(E_t+E_{\bar{t}}))/(\sqrt{s} (1-4m^2_t/s))$. Physically, when $\tau$
is non-zero, there is at least one massless parton in the final state
with finite energy. We can divide the phase space into two slices
using the theta function,
\begin{align}
\sigma^{(2)} = \sigma^{(2)}_{I} + \sigma^{(2)}_{II},
\end{align}
where $\sigma^{(2)}_{I} = \int\! \dd \sigma \theta(\delta_E - \tau)$
is the soft-virtual part,
and $\sigma^{(2)}_{II} = \int\! \dd \sigma \theta(\tau - \delta_E)$ is
the hard part,
and $\delta_E$ is the cut-off parameter. There are still phase space
singularities in both $\sigma_{I}$ and $\sigma_{II}$. However, the
phase space singularties in $\sigma_{II}$ belong to the well
understood one, because there is at most one massless parton in the
final state whose energy can approach zero. We can therefore straight-forwardly
calculate $\sigma_{II}$ using any existing NLO infrared subtraction method. On
the other hand, the soft-virtual part, $\sigma_I$, contains double un-resolved
region, whose calculation needs additional efforts. An exact
calculation for $\sigma_I$ is difficult. However, if we choose
$\delta_E$ to be small and ignore terms of $\ord{\delta_E}$, we can
calculate $\sigma_I$ using matrix elements in the soft limit, and
also expanding the phase space volume in the soft limit. Such
approximation leads to enormous simplification and makes the analytical
calculation feasible. We explain in detail the calculation of the soft-virtual
part and hard part below.

\subsection{The soft-virtual part}
\label{sec:soft}

\subsubsection{Factorization of the radiation-energy distribution}

We can write the soft-virtual part as an integral over radiation-energy distribution,
\begin{align}
  \sigma^{(2)}_I = \int^{\delta_E}_0 \! \dd \tau \frac{\dd \omega}{\dd \tau} \frac{\dd \sigma^{(2)}}{\dd \omega},
\end{align}
where $\omega$ is twice the energy of final state QCD radiations,
$\omega = 2(\sqrt{s}-E_t - E_{\bar t})$. The factor of $2$ here is
introduced by convention. 
Ignoring power suppressed terms in $\omega/m_t$, we can write the
distribution for $\dd\sigma/\dd \omega$ in small $\omega$ in a factorized form using the language
of effective theory. $\dd\sigma^{(2)}/\dd \omega$ is simply the order
$\ord{\als^2}$ corrections to this distribution.  We start from the full distribution in QCD,
\begin{align}
\frac{\dd \sigma}{\dd \omega} = \sum_{t,\bar{t},X} (2\pi)^4 \delta^{(4)}(Q - p_t
- p_{\bar t} - p_X) \delta( \omega - 2 
E(X)) L_{\mu\nu} \sum_{ij}\langle 0| J^\mu_i | t\bar{t} X\rangle
\langle t\bar{t} X | J^\nu_j | 0 \rangle \ ,
\label{eq:qcdfac}
\end{align}
where $X$ denotes gluons and light quarks in the final state. $E(X)$
denotes the energy of $X$. The lepton tensor includes only vector
contributions from virtual photon exchange,
\begin{align}
  L_{\mu\nu} = - \frac{2e^2}{s} \left( g_{\mu\nu} - \frac{2 (p^{e^+}_\mu
      p^{e^-}_\nu + p^{e^+}_\nu
      p^{e^-}_\mu)}{s} \right) \ , 
\end{align}
where $e$ is the QED coupling, and $p^{e^+}_\mu$ and $p^{e^-}_\mu$ are
the four momentum of positron and electron.
The production of top-quark pair via virtual photon exchange is
described by two QCD currents,
\begin{align}
  J^\mu_1 = -ieQ_t\bar{u}(p_t) \gamma^\mu v(p_{\bar t}) ,\qquad
J^\mu_2 = \frac{eQ_t}{2m_t} \bar{u}(p_t) \sigma^{\mu\nu} (p_t+p_{\bar
  t})_\nu v(p_{\bar t}) \ ,
\label{eq:qcdcurrent}
\end{align}
where $Q_t=2/3$ is the electric charge number of top quark, and
$\sigma^{\mu\nu} = \frac{i}{2} [\gamma^\mu,\gamma^\nu]$. Note that
Eq.~(\ref{eq:qcdfac}) is exact to leading order in electroweak
interaction, and all orders in QCD interactions. It is also an exact
distributions for $\omega$. Calculation of Eq.~(\ref{eq:qcdfac})
in perturbative QCD requires the calculation of both virtual
corrections and phase space integral. Unfortunately, exact calculation
of phase space integral is difficult beyond NLO. Certain
approximation is needed in order to proceed. Since we are only interested in
the energy distribution in the soft region, we can expand 
Eq.~(\ref{eq:qcdfac}) to leading power in $\omega/m_t$. Then
the momentum conservation delta function factorizes as
\begin{align}
  \sum_{t,\bar{t},X} (2\pi)^4 \delta^{(4)}(Q - p_t
- p_{\bar t} - p_X)  \simeq \sum_{t,\bar{t}} (2\pi)^4 \delta^{(4)}(Q - p_t
- p_{\bar t} ) \sum_X
\label{eq:phasefac}
\end{align}
in the region where $\Lambda_{QCD}
\ll \omega \ll m_t$. The physics of such factorization is that as long
as the
energy of QCD radiations is small, they can hardly change the
trajectory of heavy quark. The short-distance interaction which produces the
top-quark pair can not resolve the activities of soft QCD radiations,
therefore have tree-level like kinematics. We can describe the top quark and antitop
quark by heavy quark fields $h_{v}(y)$ and $h_{\bar v}(y)$, labeled by the velocity of the heavy
quarks, $p_t = m_t v$, $p_{\bar t}= m_t \bar{v}$.
The QCD currents  in Eq.~(\ref{eq:qcdcurrent}) can then be matched to currents in Heavy
Quark Effective Theory~(HQET),
\begin{align}
  \mathcal{J}^\mu_1 = -ieQ_t C_1(v,\bar{v})\bar{h}_{v} \gamma^\mu h_{\bar v} ,\qquad
\mathcal{J}^\mu_2 = \frac{eQ_t}{2} C_2(v,\bar{v})\bar{h}_v
\sigma^{\mu\nu} (v+\bar{v})_\nu h_{\bar v} \ , 
\label{eq:wilson}
\end{align}
where the corresponding Wilson coefficients $C_1(v,\bar{v})$ and
$C_2(v,\bar{v})$ can be obtained from the calculation of QCD form
factor for heavy quark pair production. At leading power in HQET, the
heavy quark field only interacts with gluons via eikonal interaction,
\begin{align}
  \mathcal{L}_{int}=\bar{h}_v(y) g v\cdot A_s(y) h_v(y) +
  \bar{h}_{\bar v}(y) g \bar{v}\cdot A_s(y) h_{\bar v}(y) 
\end{align}
Such eikonal interactions can be
absorbed into Wilson lines by a field
redefinition~\cite{hep-ph/0109045},
\begin{align}
  (h_v(y))^\dagger = ( h^{(0)}_v(y) )^\dagger (Y_v(y))^\dagger, \qquad
  h_{\bar v}(y) = Y_{\bar v} (y)
  h^{(0)}_{\bar v}(y) \ ,
\end{align}
where 
\begin{align}
(  Y_v(y))^\dagger = &\mathbf{P} \exp \left( ig \int^\infty_0 \dd z
  \cdot A(vz + y) \right)
\nn
\\
Y_{\bar{v}}(y) = & \overline{\mathbf{P}} \exp \left( -ig \int^\infty_0
\dd z \cdot A(\bar{v} z + y) \right)
\end{align}
are the path-ordered and anti path-ordered Wilson lines.
The decoupled heavy quark field $h^{(0)}_v(x)$ no longer interacts with
gluon, but still annihilate the top quark field.
The hadronic tensor now has a factorized form,
\begin{align}
\sum_X \delta(\omega - 2 E(X))\sum_{ij}\langle 0| \mathcal{J}^\mu_i | t\bar{t} X\rangle
\langle t\bar{t} X | \mathcal{J}^\nu_j | 0 \rangle   = H^{\mu\nu}
\sum_X
\langle 0 | Y_v^\dagger Y_{\bar v} |X\rangle \delta(\omega - 2 E(X))
\langle X| Y_{\bar v}^\dagger Y_v |0
\rangle
\label{eq:hqetfac1}
\end{align}
$H^{\mu\nu}$ is the hard function,
\begin{align}
  H^{\mu\nu} = \sum_{i,j=1}^2 \langle 0| \mathcal{J}^{(0),\mu}_i | t\bar{t} \rangle
\langle t\bar{t}  | \mathcal{J}^{(0),\nu}_j | 0 \rangle \ ,
\end{align}
and $\mathcal{J}^{(0),\mu}_i$ is the decoupled HQET current, with
$h_{v,\bar{v}}(y)$ replaced by $h_{v,\bar v}^{(0)}(y)$. Summing over
the top quark spin and color~\footnote{It should be noted that our
  formalism also allows full spin dependence for heavy quark, since the
  eikonal approximation preserves spin.}, the hard
function can be evaluated explictly,
\begin{align}
  H^{\mu\nu} =N_c  \sum_{i,j=1}^2 C_i (v,\bar v)C_j^*(v,\bar v) h^{\mu\nu}_{ij},
\end{align}
with
\begin{align}
  h^{\mu\nu}_{11} =& e^2 Q^2_t \left( - 2 s g^{\mu\nu} +
    4(p^\mu_t p^\nu_{\bar t} + p^\nu_t p^\mu_{\bar t}) \right)\ ,
\\
h^{\mu\nu}_{12} = &  h^{\mu\nu}_{21} = 2 e^2 Q^2_t \left( -s g^{\mu\nu}
  + (p^\mu_t + p^\mu_{\bar t})(p^\nu_t + p^\nu_{\bar{t}}) \right)\ ,
\\
h^{\mu\nu}_{22} = & e^2 Q^2_t \left( -2 s g^{\mu\nu} + \left( 2 -
    \frac{s}{2m^2_t}\right) (p^\mu_t p^\nu_{\bar t} + p^\nu_t
  p^\mu_{\bar t} ) + \left( 2 + \frac{s}{2 m^2_t} \right) (p^\mu_t
  p^\nu_t + p^\mu_{\bar t} p^\nu_{\bar t}) \right)
\end{align}
$N_c=3$ is the number of color in QCD.
The matrix element of the Wilson lines defines the soft function for
$t\bar{t}$ production,
\begin{align}
  S = \frac{1}{N_c} \sum_X
\langle 0 | Y_v^\dagger Y_{\bar v} |X\rangle \delta(\omega - 2 E(X))
\langle X| Y_{\bar v}^\dagger Y_v |0
\rangle \ ,
\end{align}
The summation is over all possible QCD final states. We have chosen
the normalization such that at LO the soft function is $\delta(\omega)$.
The calculation for soft function is much easier than the exact phase space
integral, thanks to the eikonal approximation. 

We can now write down a factorized formula for the radiation-energy
distribution in top-quark pair production,
\begin{align}
 \frac{\dd \sigma_{\mathrm{s.v.}} }{\dd \omega} =\frac{1}{8s} \int\! \frac{d^3 p_t}{2E_t(2\pi)^3}
  \int\! \frac{d^3 p_{\bar t}}{2E_{\bar t}(2\pi)^3} (2\pi)^4 \delta^{(4)}
  (Q - p_t -p_{\bar t}) L_{\mu\nu} H^{\mu\nu} S(x, \omega)  \ ,
\label{eq:softdis1}
\end{align}
where we have also included the initial dtate flux and spin average factor. The variable $x$ is defined as
\begin{align}
  x = \frac{ 1 - \sqrt{ 1  - \frac{4 m^2_t}{s} }}{ 1 + \sqrt{ 1  - \frac{4 m^2_t}{s} }}
\end{align}
For fixed $m_t$, $x\to 0$ is the high energy limit, while $x\to 1$ is
the threshold limit.
Eq.~(\ref{eq:softdis1}) is only valid at leading power in $\omega$. The soft function is fully
differential in the top and antitop momentum, but inclusive in the
QCD radiations. This is not a problem as we will use this formula only in the limit of small
$\omega$, where the QCD radiations can not be resolved by any
reasonable experimental measurement.

The phase space integral in 
Eq.~(\ref{eq:softdis1}) becomes trivial. Integrating out the azimuthal
angle dependence of top quark, we obtain
\begin{align}
  \frac{\dd^2 \sigma_{\mathrm{s.v.}} }{\dd \omega\, \dd \cos\theta_t}  =
  \mathcal{H}(\cos\theta_t, m_t, s)S(x, \omega)  \ ,
\end{align}
where 
\begin{align}
  \mathcal{H}(\cos\theta_t, m_t, s)  =& \frac{1}{8s} \int\! \frac{d^3 p_t}{2E_t(2\pi)^3}
  \int\! \frac{d^3 p_{\bar t}}{2E_{\bar t}(2\pi)^3}
\nn
\\
&\times (2\pi)^4 \delta^{(4)}
  (Q - p_t -p_{\bar t}) \delta \left(\cos\theta_t - \frac{\mb{p}_t
      \cdot \mb{p}_e }{|\mb{p}_t||\mb{p}_e|} \right)  L_{\mu\nu} H^{\mu\nu} 
\end{align}
The soft function is a distribution in $\omega$. It is often convenient
to perform a Laplace transformation,
\begin{align}
  \frac{\dd^2 \tilde \sigma_{\mathrm{s.v.}} }{d\kappa \, \dd \cos\theta_t} = &\int^\infty_0 \! \dd \omega
  \exp\left( - \frac{\omega}{e^{\gamma_E} \kappa} \right) \frac{\dd^2
    \sigma_{\mathrm{s.v.}}}{\dd \omega \, \dd \cos\theta_t} 
\nn
\\
\equiv & \mathcal{H}(\cos\theta_t , m_t ,s)\tilde s(x, L_\kappa)  \ ,
\end{align}
where $L_\kappa = \ln (\kappa/\mu)$. The renormalized soft function
depends on $\kappa$ only through terms of the form $L^n_\kappa$, where
$n$ is a positive integer. It is therefore possible to invert the
Laplace transformation in close form~\cite{hep-ph/0607228},
\begin{align}
  \frac{\dd^2 \sigma_{\mathrm{s.v.}} }{\dd \tau\, \dd \cos\theta_t} = \mathcal{H}(\cos\theta_t , m_t ,s) \lim_{\eta \to
    0} \left[  \tilde{s} \left(\partial_\eta
   \right)\left( \frac{\mu}{\sqrt{s}(1-4m^2_t/s)}\right)^{\eta}  \frac{1}{\tau^{1+\eta}}
  \frac{\exp(-\gamma_E \eta)}{\Gamma(\eta)} \right] \ ,
\label{eq:master}
\end{align}
where we recall that $\tau = \omega/(\sqrt{s}(1-4m^2_t/s))$.
 Eq.~(\ref{eq:master}) is interpreted as first expanding in $\eta$ as
 a taylor series within the
 square bracket, using the well-known plus-distribution expansion
 \begin{align}
   \frac{1}{\tau^{1+\eta}}  = - \frac{\delta(\tau)}{\eta} +
   \frac{1}{[\tau]_+} - \eta \left[\frac{\ln \tau}{\tau}\right]_+ +
   \mathcal{O}(\eta^2) \ ,
 \end{align}
then taking the $\eta \to 0$ limit. 

\subsubsection{Hard function from QCD heavy quark form factor}

The Wilson coefficients defined in Eq.~(\ref{eq:wilson}) can be
obtained from the QCD heavy quark form factor. The latter has been
computed for the vector contributions, axial contributions, and
anomaly contributions by Bernreuther
et.al~\cite{hep-ph/0406046,hep-ph/0412259,hep-ph/0504190}.  The vector
contributions have been computed independently later in ref.~\cite{0905.1137},
confirming previous results.

In ref.~\cite{hep-ph/0406046}, the vector contributions to heavy quark
form factor are given to two loops in QCD. The results are
expressed in terms of two dimensionless scalar form factors, $\hat{F}_1(x)$ and
$\hat{F}_2(x)$,
\begin{align}
  -i e Q_t \bar{u}(p_t) \left( \hat{F}_1(x) \gamma^\mu + \frac{1}{2 m_t} \hat{F}_2(x) i
    \sigma^{\mu\nu} (p^t_\nu + p^{\bar t}_\nu) \right) v(p_{\bar t}) 
\end{align}
Here the scalar form factors are related to those computed in Eq.~(57)
and (58) of
ref.~\cite{hep-ph/0406046} by an additional renormalization,
\begin{align}
  \hat{F}_i(x,\als^{N_l}) = F_i (x,\als^{N_f}(\als^{N_l})) \ , \qquad i=1,2 \ ,
\label{eq:finiteren}
\end{align}
where
\begin{align}
\als^{N_f} (\als^{N_l}) = \als^{N_l} \left[1+\frac{8}{3}T_RN_h \frac{\als^{N_l}}{4 \pi} \left( -
    \frac{1}{2} L_H + \ep \left( \frac{L^2_H}{4} + \frac{1}{24}\pi^2
    \right) \right) +\frac{\ep}{12}  \left(\frac{\als^{N_l}}{4 \pi} \right)\beta^{(N_f)}_0 \pi^2
\right] \ ,
\label{eq:finitediff}
\end{align}
with $N_f = N_l + N_h$. $N_l=5$ is the number of light quark
flavor, and $N_h=1$ is the number of heavy quark flavor. $T_R = 1/2$
in QCD, $L_H = \ln( m^2_t/\mu^2) $.  The QCD beta function for $N_f$
quark flavor is given by
\begin{align}
  \beta^{(N_f)}_0 = \frac{11}{3} C_A - \frac{4}{3} T_R (N_l + N_h) 
\end{align}
where $C_A = 3$ in QCD. Unless otherwise specified, we will denote $\als^{N_l}$
as $\als$ below. Note that $\hat{F}_i(x)$ and $F_i(x)$ only
differ starting from two loops. The origin for such difference is that in ref.~\cite{hep-ph/0406046}, the
renormalization of strong coupling is performed in
$\overline{\mathrm{MS}}$ scheme, running with $N_f$ flavors. Also the
authors of ref.~\cite{hep-ph/0406046} include a factor $\Gamma(1+\ep)
\exp(\ep \gamma_E)$ in the coupling renormalization, where $\Gamma(z)$
is Euler's Gamma function, and $\gamma_E = 0.577216 \dots$. However, we
choose to perform the calculation with $\als$ running with $N_l$
flavors, and also without the additional factor $\Gamma(1+\ep)
\exp(\ep \gamma_E)$. The decoupling of heavy quark flavor is realized
by the second
term on the RHS of Eq.~(\ref{eq:finitediff})~\cite{hep-ph/9411260,hep-ph/9706430,hep-ph/9708255,hep-ph/0004189}, while the third factor
gets rid of the additional factor $\Gamma(1+\ep)
\exp(\ep \gamma_E)$ through to $\ord{\als^2}$~\cite{hep-ph/0612149}.

The scalar form factors are functions of $x$.
 Writing them as an expansion
in $a_s= \als(\mu)/(4\pi) $, 
\begin{align}
  \hat F_i(x) = \hat F^{(0l)}_i(x) + a_s\hat  F^{(1l)}_i(x) + a^2_s\hat F^{(2l)}_i(x) +
  \ord{a^3_s} \ ,
\end{align}
we have at LO in QCD
\begin{align}
\hat  F^{(0l)}_1 (x) = 1, \qquad \hat F^{(0l)}_2 (x) = 0
\end{align}
Using the additional renormalization relation in
Eq.~(\ref{eq:finiteren}), the one-loop and two-loop form factors can
be read off from ref.~\cite{hep-ph/0406046}.  These form factors
are UV finite but IR divergent. To calculate the Wilson coefficients defined in
Eq.~(\ref{eq:wilson}), one needs to calculate the form factors in the
effective theory. The wilson coefficients are simply the differences of
the form factor in QCD and the form factor in effective theory. In dimensional regularization with external state
onshell, the form factors in the effective theory at one loop and
beyond vanish because they invlove only scaleless integral. Since the
IR divergences in the QCD calculation and effective theory
calculation must match, it implies that the UV divergences in the
effective theory calculation are exactly the negative of the IR
divergence in the QCD calculation. Therefore, renormalization of the
UV divergences in the effective theory is simply amount to performing an IR subtraction to the
form factor in QCD,
\begin{align}
  C_i(x) =  \lim_{\ep \to 0}\left[ Z_{H,i} \hat{F}_i(x)\right] \ , \qquad i=1,2 \ ,
\end{align}
where the IR subtraction factor is defined such that $C_i(x)$ is order
by order finite, i.e., $Z_{H,i}$ absorbs \emph{only} the $\ep$ poles in
$\hat{F}_i(x)$. For the convenience of reader, we give below the
explicit expression for $C_i(x)$ at one loop, as derived from the QCD form factors in
ref.~\cite{hep-ph/0406046}. We have checked that using
ref.~\cite{0905.1137}, we get the same Wilson coefficients.

The one-loop Wilson coefficients are
\begin{align}
  C^{(1l)}_1(x) = & C_F \left[L_H \left(\left(\frac{2}{x+1}-\frac{2}{x-1}-2\right)
   H(0,x)+2\right)+\left(-\frac{4}{x+1}+\frac{2}{x-1}+3\right) H(0,x)
\right.
\nn
\\
&
\left.
+\left(-\frac{2}{x+1}+\frac{2}{x-1}+2\right)
   H(0,0,x)+\left(-\frac{4}{x+1}+\frac{4}{x-1}+4\right) H(1,0,x)
\right.
\nn
\\
&
\left.
-\frac{8 \zeta_2}{x-1}+\frac{8 \zeta_2}{x+1}-4 (2
   \zeta_2+1)\right] + i \pi C_F
 \left[L_H \left(\frac{2}{x+1}-\frac{2}{x-1}-2\right)
\right.
\nn
\\
&
\left.
+\left(-\frac{2}{x+1}+\frac{2}{x-1}+2\right) H(0,x)
+\left(-\frac{4}{x+1}+\frac{4}{x-1}+4\right)
   H(1,x)
\right.
\nn
\\
&
\left.
+\frac{2}{x-1}-\frac{4}{x+1}+3\right]
\\
C^{(1l)}_2(x) = & 2 C_F \left( -\frac{1}{x-1}-\frac{1}{x+1} \right)
(H(0,x)+ i \pi  ) \ ,
\end{align}
where $C_F = 4/3$ in QCD. The imaginary part in the Wilson coefficients results from analytical continuation of the form factors from spacelike to timelike kinematics. The function $H(\vec{w},x)$ is harmonic
polylogarithm~(HPL) introduced in ref.~\cite{hep-ph/9905237}.  We use \texttt{hplog}~\cite{hep-ph/0107173} for the numerical
calculation of HPLs in this work.
The \texttt{Mathematica} file for the two-loop Wilson coefficients can be found in the arXiv submission
of this paper.

\subsubsection{Perturbative expansion of the radiation-energy distribution through to NNLO}

To expand the equation for radiation-energy distribution in
Eq.~(\ref{eq:master}) in $\als$, we also need the soft function to
NNLO, which have been computed only recently~\cite{vMSZ}. The Laplace
transformed soft function has the generic form 
\begin{eqnarray}
  \ws(x, L_\kappa) &=& 1 + a_s \Big(L_\kappa\, \gamma^s_0(x) +
  c_1(x) \Big)+ a_s^2 \bigg[ L_\kappa^2 \left(\frac{1}{2} \Big(\gamma^s_0(x)\Big)^2 -
\beta_0 \gamma^s_0(x)\right)
\nn
\\
&& +
L_\kappa \left(c_1(x)\Big(\gamma^s_0(x) - 2\beta_0\Big) 
+ \gamma^s_1(x)\right) + c_2(x) \bigg] 
+ \mathcal{O}\left(a^3_s\right).
\label{eq:rens}
\end{eqnarray}
through to $\ord{\als^2}$, 
where $\beta_0$ is the LO QCD beta function with $N_l$ light flavour only,
\begin{equation}
\beta_0  = \frac{11}{3}C_A - \frac{4}{3} T_R N_l\ ,
\end{equation} and
$\gamma_0^s(x)$ and $\gamma_1^s(x)$ are the well-known cusp anomalous
dimension~\cite{NUPHA.B283.342,0903.2561,0904.1021}. We reproduce them here for the sake of
completeness
\begin{eqnarray}
&&\gamma^s_0(x) = -8 C_F \left[1+\frac{1+x^2}{1-x^2}H(0,x)\right]
\label{eq:anomalous1}
\\
&&\gamma^s_1(x) = \frac{160}{9} C_F N_l T_R \left[1+\frac{1+x^2}{1-x^2}H(0,x)\right]
\label{eq:anomalous2}
\\ &&
+ C_A C_F \left[
-\frac{392}{9}+16 \zeta_2 \frac{1 + 9 x^2}{1-x^2}+16\frac{\left(1+x^2\right)^2}{\left(1-x^2\right)^2} \Big(2 H(0,x) \Big(H(0,-1,x)-H(0,1,x)\Big)\nn
\right.\\&&\left.
-4 H(0,0,-1,x)+4 H(0,0,1,x)-\zeta (3)\Big)+32\frac{1+x^2}{1-x^2} \bigg(H (0,x) \left( H(-1,x)-H(1,x)-\frac{67}{36}\right)\nn
\right.\\&&\left.
-H(0,-1,x)+H(0,1,x)\bigg)-\frac{32}{3}\frac{x^2(1+x^2)}{\left(1-x^2\right)^2}H^3(0,x)-\frac{32 x^2}{1-x^2} H^2(0,x)\nn
\right.\\&&\left.
+ 16 \zeta_2 \frac{\left(1+x^2\right)\left(1+9 x^2\right)}{\left(1-x^2\right)^2}H(0,x)
\right].\nn
\end{eqnarray}
The soft function is largely fixed by the renormalization group
equation it obeys~\cite{NUPHA.B283.342}. The genuine two-loop
corrections to the soft function are summarized by the scalar function
$c_2(x)$, which is first computed in ref.~\cite{vMSZ}~\footnote{Note that results presented in ref.~\cite{vMSZ} are
  given in terms of generalized polylogarithms, $G(\cdots; x)$, with
  weight alphabet drawn from $\{-1,0,1\}$. They are related to HPLs by
a simple relation, $G(\vec{w};x) = (-1)^{n_1} H(\vec{w},x)$, where
$n_1$ is the number of occurence of alphabet $1$ in the weight vector $\vec{w}$.}. With all these results at hand, we can write down the
radiation-energy distribution through to NNLO, up to power-correction terms in
$\tau$. Writing $\mathcal{H}(\cos\theta_t,m_t,s)$ as an expansion in
$a_s$, $\mathcal{H}(\cos\theta_t,m_t,s) = \mathcal{H}_0 +
a_s\mathcal{H}_1 + a_s^2 \mathcal{H}_2 + \cdots$,  the results are
\begin{align}
  \frac{\dd^2 \sigma^{(1l)}_{\mathrm{s.v.}} }{\dd \tau \, \dd \cos\theta_t} = & \left[
   ( c_1(x) + L_H \gamma^s_0) \mathcal{H}_0 + \mathcal{H}_1 \right]
 \delta(\tau) + 2
  \gamma^s_1(x) \mathcal{H}_0 \frac{1}{[\tau]_+} 
\\
 \frac{\dd^2 \sigma^{(2l)}_{\mathrm{s.v.}} }{\dd \tau \, \dd \cos\theta_t} =
 &\Big[ \frac{1}{2}  \mathcal{H}_0 L_H^2 \gamma^s_0 \left(\gamma^s_0-\beta
   _0\right) + L_H \left(\mathcal{H}_0 \left(c_1(x)
       \gamma^s_0-\beta_0 c_1(x) +\gamma^s_1\right)+\mathcal{H}_1
     \gamma^s_0 \right)
\nn
\\
&
+\mathcal{H}_0 \left(c_2(x)+\frac{1}{3} \pi ^2 \beta_0
  \gamma^s_0-\frac{1}{3} \pi ^2
  \left(\gamma^s_0\right)^2\right)+\mathcal{H}_1
c_1(x)+\mathcal{H}_2\Big] \delta(\tau) 
\nn
\\
&
+ \Big[ \mathcal{H}_0
   L_H \left(2 \left(\gamma^s_0\right)^2+ \mathcal{H}_0 \left(2 c_1(x)
       \gamma^s_0-2 \beta_0 c_1(x)+2 \gamma^s_1\right)-2 \beta_0
     \gamma^s_0\right)+2 \mathcal{H}_1 \gamma^s_0\Big] \frac{1}{[\tau]_+}
\nn
\\
&
-4
    \mathcal{H}_0 \gamma^s_0 \left(\beta_0-\gamma^s_0\right) \left[\frac{\ln\tau}{\tau}\right]_+
\end{align}
This is the main results for the soft-virtual part.

\subsection{The hard part}
\label{sec:hard}
The hard part $\sigma^{(2)}_{II}$ consists of the real-virtual corrections, $e^+e^-\to t\bar t g$
at one loop, and the double-real corrections, $e^+e^-\to t\bar t gg(q\bar q)$ at tree level.
As mentioned above, the infrared divergences in this part only
involve single unresolved limit, thus can be extracted using standard
NLO subtraction technique. In this paper we employ the massive version
of dipole subtraction method~\cite{Catani:2002hc}. The one-loop
real-virtual calculation is carried out by the automated
program \texttt{GoSam2.0}~\cite{Cullen:2014yla} with loop integral reductions from
\texttt{Ninja}~\cite{Mastrolia:2012bu,Peraro:2014cba} and scalar integrals from \texttt{OneLOop}~\cite{vanHameren:2009dr,vanHameren:2010cp}.
Since $\sigma^{(2)}_{II}$ is IR finite, it can be compared directly to
the NLO QCD calculation of $e^+e^-\to Q\bar Q g$, {\it e.g.},
ref.~\cite{hep-ph/9705295}, and shows very good agreements.

Once the soft-virtual part and hard part are known, the full corrections are
simply the sum of them. The soft-virtual part has born kinematics in the final
state, since the QCD radiations are soft and have been integrated out. Its numerical
implementation is therefore trivial. The hard part is nothing but the
usual NLO QCD corrections to the process $e^+e^-\to t\bar{t} g$, as
described above. We believe this is the most important advantage of
phase-space slicing method, because its numerical implementation is no
more difficult than a typical NLO calculation.

However, the drawback of phase-slicing method is also clear. In
principle, the sum of the soft-virtual part and hard part is independent of
the arbitrary cut-off parameter $\delta_E$ in the limit of $\delta_E
\to 0$. Furthermore, since we will approximate the kinematics of the soft
part as born kinematics in our numerical calculation, $\delta_E$ needs to be small for such
approximation to hold.  In realistic calculation, such a limit can never be reached in
the hard part. Nevertheless, our formalism is exact in the hard part,
and include all the leading singular
dependence of $\delta_E$ in the soft-virtual part, such that the sum only
depends mildly on $\delta_E$. To estimate the form of the subleading term
missing in the soft-virtual part, we note that an exact $\tau$ distribution in
small $\tau$ should have the following form
\begin{align}
  \frac{\dd \sigma^{(2)}}{\dd \tau} = A(x) \left[\frac{\ln\tau}{\tau}\right]_+ +
  \frac{B(x)}{[\tau]_+}  + C(x) \delta(\tau) + D(x) \ln\tau + \text{subleading terms}
\end{align}
Our calculation includes exact results for the first three
coefficients, $A(x)$, $B(x)$ and $C(x)$, but not $D(x)$. Integrating
over the fourth term over $\tau$ gives
\begin{align}
D(x)  \int^{\delta_E}_0 \! \dd \tau \ln\tau \simeq D(x) \delta_E \ln
\delta_E + \text{subleading terms in } \tau
\end{align}
We therefore expect the leading missing $\delta_E$ dependence in the
sum of the soft-virtual part and hard part is proportional to
$\delta_E\ln\delta_E$ at NNLO. To minimize the impact of such
contributions, we have to choose very small cut-off parameter
$\delta_E$. This is not a problem for the soft-virtual part, as $\delta_E$
dependence there is analytical. For the hard part, choosing extremely
small $\delta_E$ leads to finite but very large corrections, comparing
to the corrections to the sum. Thus there has to be delicate cancelation
of large corrections between the soft-virtual part and hard part. A possible
improvement would be including also the subleading terms $D(x)
\ln\tau$ in the calculation. Such ``next-to-eikonal corrections'' have
been considered before in Drell-Yan production through to
NNLO~\cite{1007.0624,1010.1860}. It would be interesting to calculate
$D(x)$ along the same line.

\section{N\lowercase{umerical results}}
\label{sec:num}

We present our numeric results in this section. As mentioned before, we use two-loop running of the QCD coupling
constants with $N_l=5$ active quark flavors
and $\alpha_s(M_Z)=0.118$. We choose the $G_F$ parametrization scheme~\cite{Denner:1990ns} for the EW
couplings with $M_W=80.385\,{\rm GeV}$, $M_Z=91.1876\,{\rm GeV}$, $M_{t}=173\,{\rm GeV}$,
and $G_F=1.166379\times 10^{-5}\,{\rm GeV}^{-2}$~\cite{Beringer:1900zz}. The renormalization scale is
set to the center of mass energy $\sqrt s$ unless otherwise specified.

The production cross sections due to virtual photon exchange through to NNLO in QCD can be expressed as
\begin{equation}\label{eq:kfa}
\sigma_{NNLO, \gamma}=\sigma_{LO, \gamma}\left(1+\Delta^{(1), \gamma}+\Delta^{(2),
\gamma}\right),
\end{equation}
where $\Delta^{(1,2), \gamma}$ denote respectively the $\ord{\als}$ and
$\ord{\als^2}$ QCD corrections. The $\ord{\als^2}$ corrections
$\Delta^{(2), \gamma}$ can be further decomposed according to color factors, i.e., the
Abelian contributions, the non-Abelian contributions, the light-fermionic contributions, and the
heavy-fermionic contributions. Alternative notation used in~\cite{hep-ph/9712222,hep-ph/9710413,hep-ph/9704222} follows
\begin{equation}
\sigma_{NNLO, \gamma}=\sigma_{\mu^+\mu^-, \gamma}\left(R^{(0)}+\frac{\alpha_s(\mu^2)}{\pi}C_F R^{(1)}
+\left(\frac{\alpha_s(\mu^2)}{\pi}
\right)^2R^{(2)}\right),
\end{equation}
with $\sigma_{\mu^+\mu^-,\gamma}$ be the cross section of muon pair production, and
\begin{equation}\label{eq:cdec}
R^{(2)}=C_F^2R^{(2)}_A+C_AC_FR^{(2)}_{NA}+C_FT_RN_lR^{(2)}_{lF}+C_FT_RR^{(2)}_{hF},
\end{equation}
depends only on $r=2m_t/\sqrt{s}$. The four contributions in Eq.~(\ref{eq:cdec})
are denoted by ``$C_F$'', ``$C_A$'', ``$N_l$'', and ``$N_h$'' respectively in the
following figures and discussions. Analytical results for $R^{(2)}$ are presented
for production near threshold~\cite{hep-ph/9712222,hep-ph/9712302} or in the
high energy expansions~\cite{hep-ph/9710413,hep-ph/9704222}
with which we compare our numerical results.

\subsection{Inclusive cross sections\label{sec:inc}}
\begin{figure}[H]
  \begin{center}
  \includegraphics[width=0.4\textwidth]{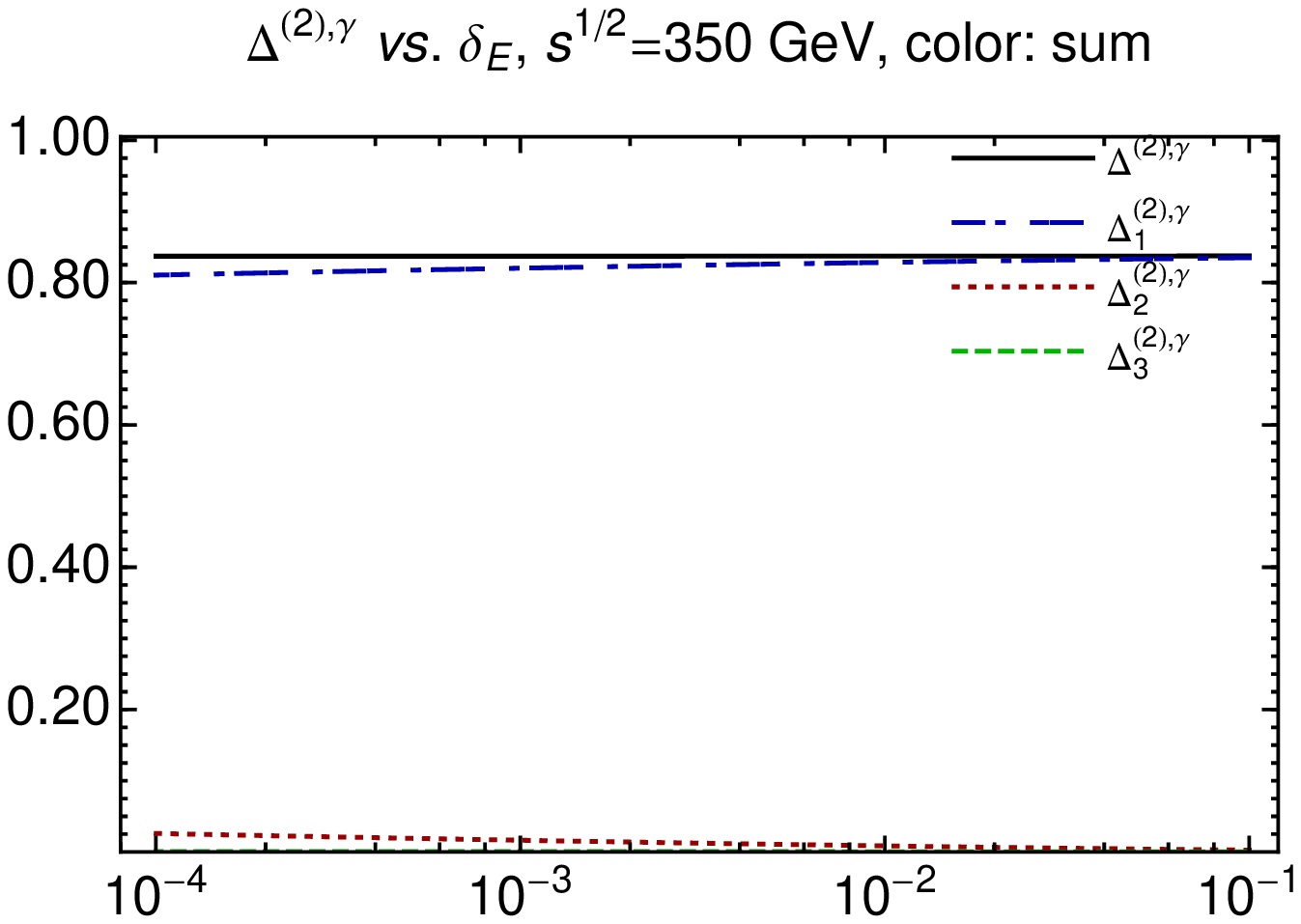}
  \includegraphics[width=0.4\textwidth]{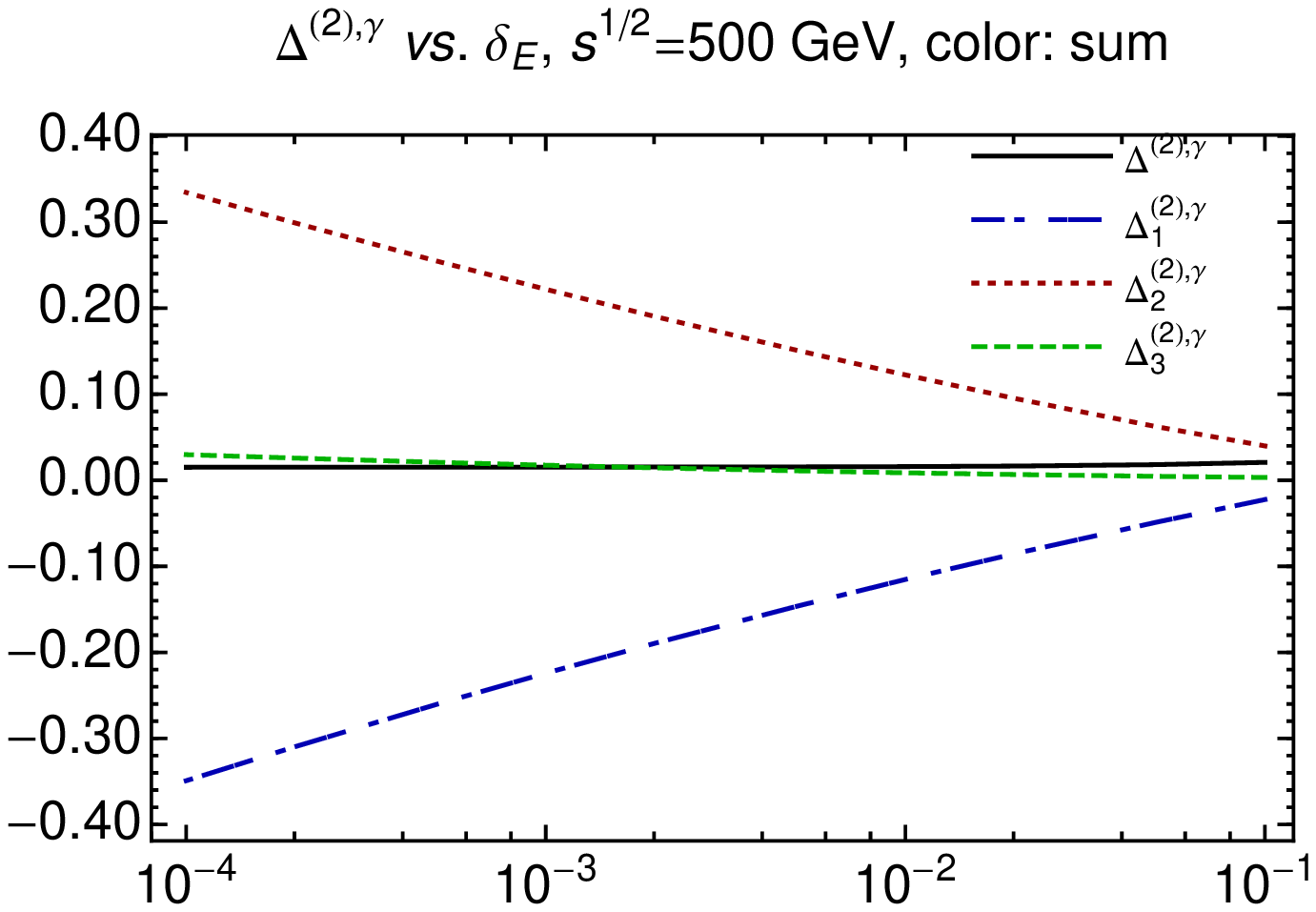}\\
  \includegraphics[width=0.4\textwidth]{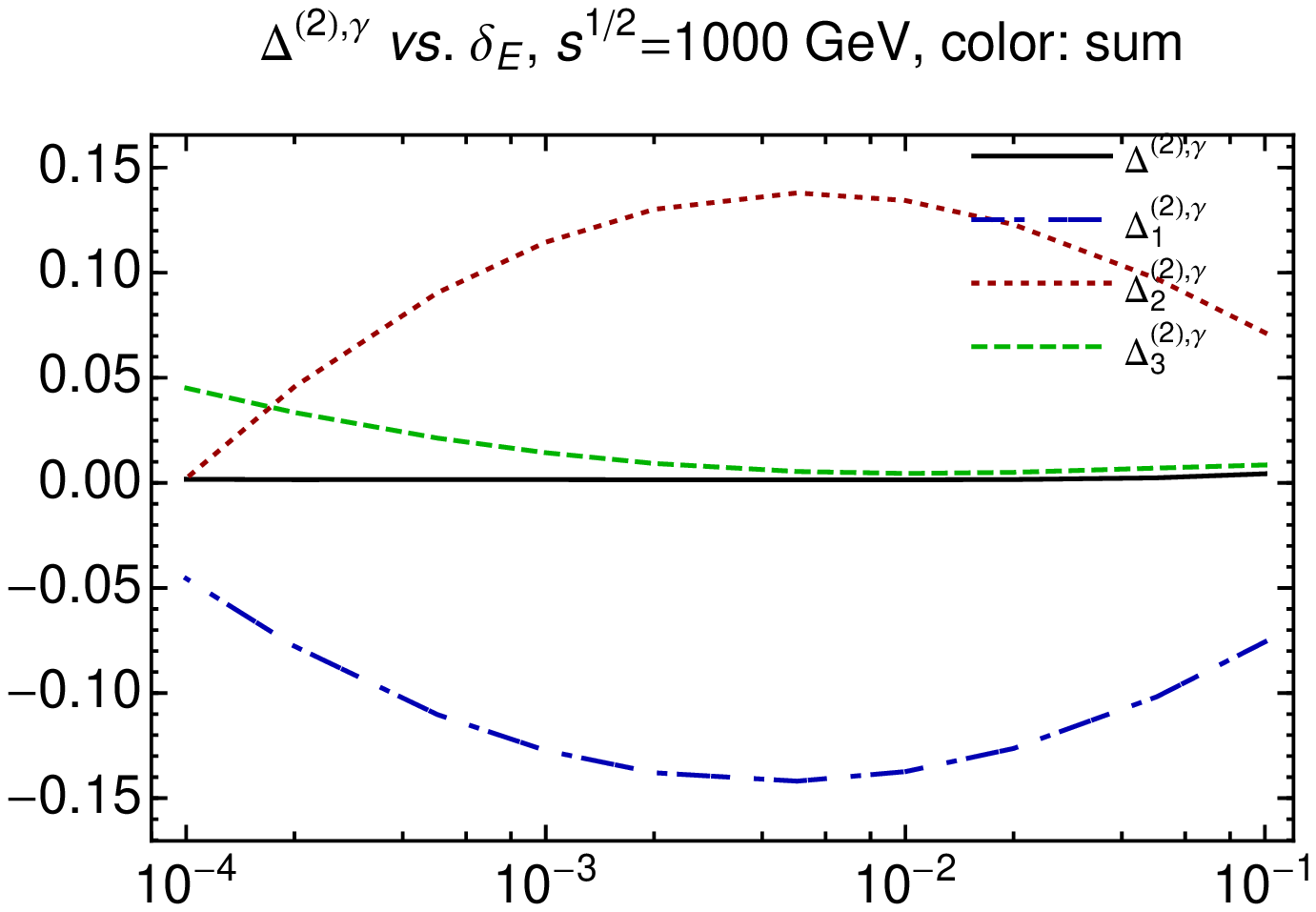}
  \end{center}
  \vspace{-1ex}
  \caption{\label{fig:des1}
  Dependence of separate contributions to $\Delta^{(2),\gamma}$ with
  full colors on the cut-off for different collision energies.}
\end{figure}

\begin{figure}[H]
  \begin{center}
  \includegraphics[width=0.4\textwidth]{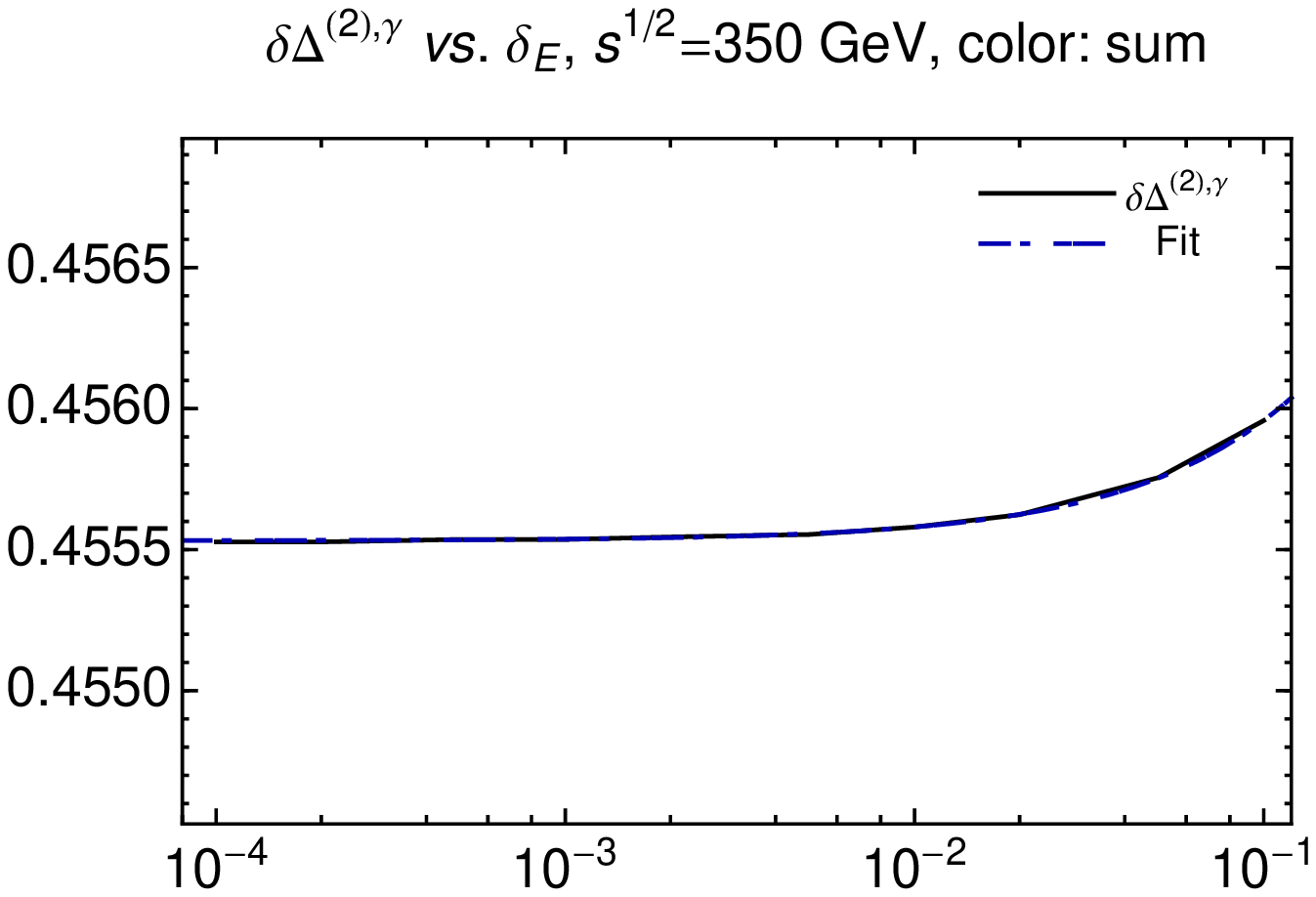}
  \includegraphics[width=0.4\textwidth]{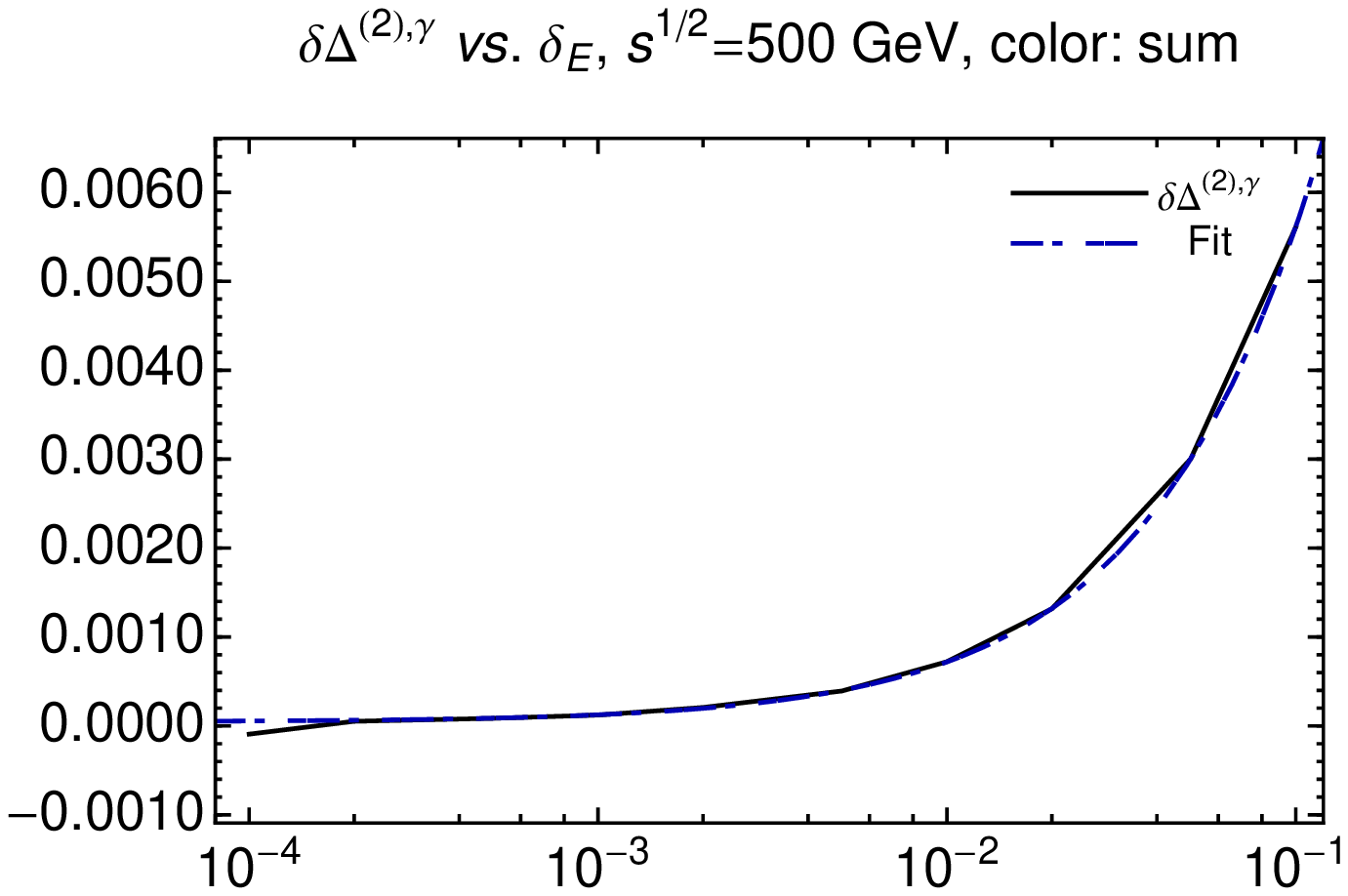}\\
  \includegraphics[width=0.4\textwidth]{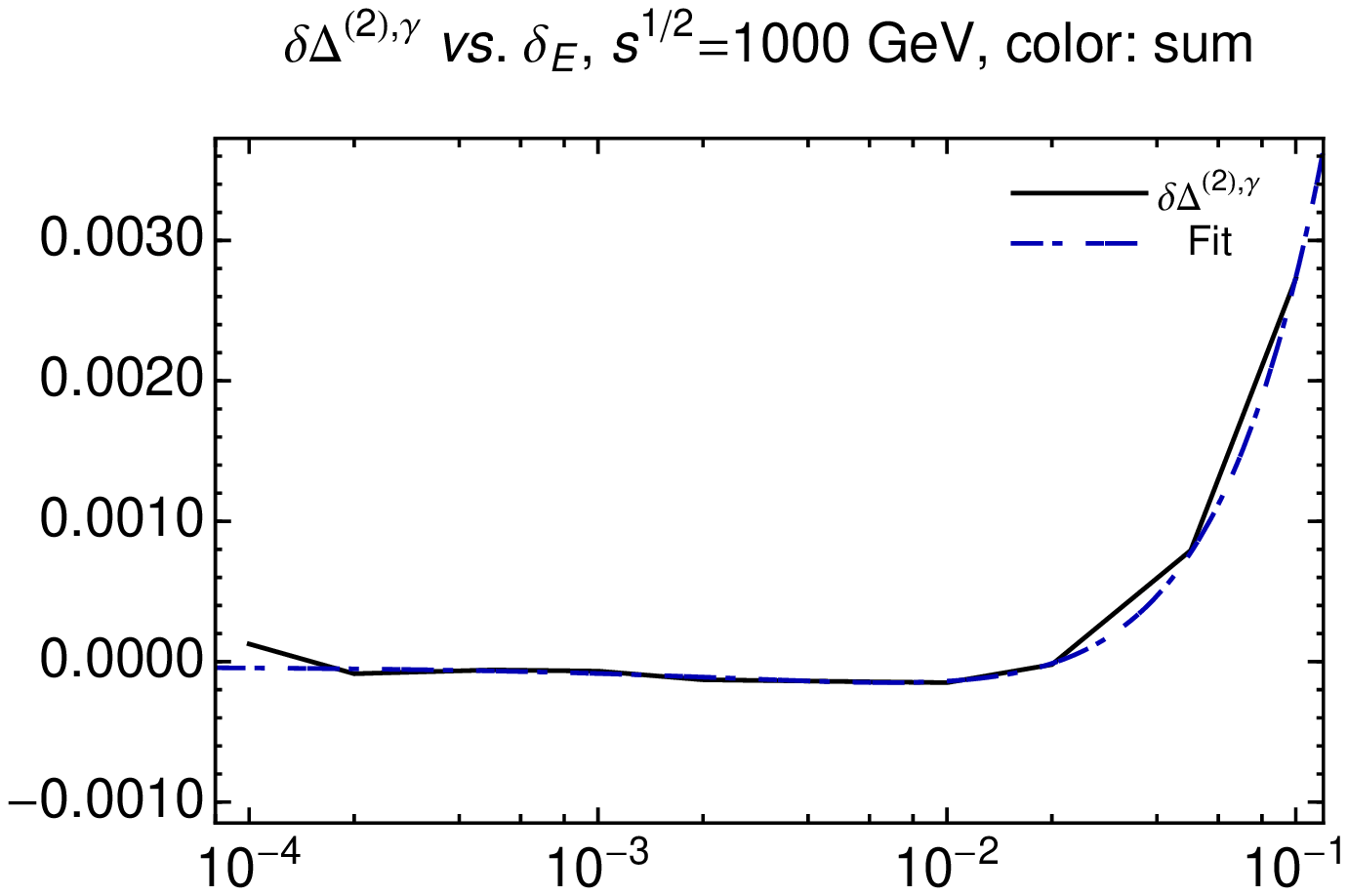}
  \end{center}
  \vspace{-1ex}
  \caption{\label{fig:des2}
  Dependence of $\delta \Delta^{(2),\gamma}$ with full colors on the
  cut-off and the fitted curves for different collision energies.}
\end{figure}

As usual in phase-space slicing method, $\Delta^{(2),\gamma}$ depends only weakly on
the cut-off parameter $\delta_E$ and approaches the genuine $\ordb$ corrections when $\delta_E$
is small enough. Fig.~\ref{fig:des1} shows $\Delta^{(2), \gamma}$ as functions of $\delta_E$ for
different collision energies. For each of the energy choices, $\Delta^{(2), \gamma}$ receives
contributions from below the cut-off $\Delta^{(2), \gamma}_1$ (soft-virtual part), and above the cut-off
$\Delta^{(2), \gamma}_{2/3}$ (hard parts). Each of the three parts depends strongly on $\delta_E$ with
variations as large as 30\% for example for $\sqrt s=500 \,{\rm
  GeV}$. However, their sum,
$\Delta^{(2), \gamma}$ remains almost unchanged when $\delta_E$ varies between $10^{-2}$ and $10^{-4}$ as
demonstrated in Fig.~\ref{fig:des1}. For production near the threshold, e.g., $\sqrt s=350\,{\rm GeV}$,
the dominant contribution to the $\ordb$ corrections is from the two-loop virtual corrections
as included in $\Delta^{(2),\gamma}_1$.
The remaining dependences of $\Delta^{(2),\gamma}$ on
$\delta_E$ are further plotted in Fig.~\ref{fig:des2}. Here in $\delta\Delta^{(2),\gamma}$ we have
subtracted the high energy expansion results~\cite{hep-ph/9710413,hep-ph/9704222} from our numerical results for
comparison. The solid lines are scattering plots and the dashed lines are fitted curves assuming
$\delta\Delta^{(2),\gamma}=f_0+f_1\delta_E\ln\delta_E+f_2\delta_E$, where $f_i$
are constants independent of $\delta_E$. The fitted coefficients are $f_{0,1,2}=\{0.4555, -0.00025, 0.0037\}$,
$\{0.00005, -0.0050, 0.044\}$, and $\{-0.00006, 0.026, 0.070\}$ for the three collision
energies respectively. Note that the $f_0$ term represents difference of our numerical
results in the limit of $\delta_E\to 0$ (genuine $\ordb$ corrections) with the high energy
expansion results. $f_{1}$ and $f_{2}$ terms are the systematic errors due to
finite $\delta_E$ choices. Assuming $\delta_E=2\times10^{-4}$, the $f_1$ and $f_2$ terms
add up to less than $10^{-4}$ for above collision energies. Thus choosing $\delta_E=2\times 10^{-4}$
should be sufficient for a realistic calculation. The smalless of $f_0$ for $\sqrt s=500$ and
$1000$ GeV indicates a very good agreements of our numerical results with the high energy
expansion ones. %Similar results for separate color contributions of $\delta\Delta^{(2),\gamma}$
%are given in Figs.~\ref{fig:des3}-\ref{fig:des5}.

%\begin{figure}[h]
%  \begin{center}
%  \includegraphics[width=0.32\textwidth]{eps/desm350cfcom2.eps}
%  \includegraphics[width=0.32\textwidth]{eps/desm500cfcom2.eps}
%  \includegraphics[width=0.32\textwidth]{eps/desm1000cfcom2.eps}
%  \end{center}
%  \vspace{-1ex}
%  \caption{\label{fig:des3}
%  Dependence of $\delta \Delta^{(2),\gamma}$ with Abelian contributions on the
%  cut-off and the fitted curves for different collision energies.}
%\end{figure}
%
%\begin{figure}[h]
%  \begin{center}
%  \includegraphics[width=0.32\textwidth]{eps/desm350cacom2.eps}
%  \includegraphics[width=0.32\textwidth]{eps/desm500cacom2.eps}
%  \includegraphics[width=0.32\textwidth]{eps/desm1000cacom2.eps}
%  \end{center}
%  \vspace{-1ex}
%  \caption{\label{fig:des4}
%  Dependence of $\delta \Delta^{(2),\gamma}$ with non-Abelian contributions on the
%  cut-off and the fitted curves for different collision energies.}
%\end{figure}
%
%\begin{figure}[h]
%  \begin{center}
%  \includegraphics[width=0.32\textwidth]{eps/desm350nlcom2.eps}
%  \includegraphics[width=0.32\textwidth]{eps/desm500nlcom2.eps}
%  \includegraphics[width=0.32\textwidth]{eps/desm1000nlcom2.eps}
%  \end{center}
%  \vspace{-1ex}
%  \caption{\label{fig:des5}
%  Dependence of $\delta \Delta^{(2),\gamma}$ with light-fermionic contributions on the
%  cut-off and the fitted curves for different collision energies.}
%\end{figure}

\begin{figure}[h]
  \begin{center}
  \includegraphics[width=0.4\textwidth]{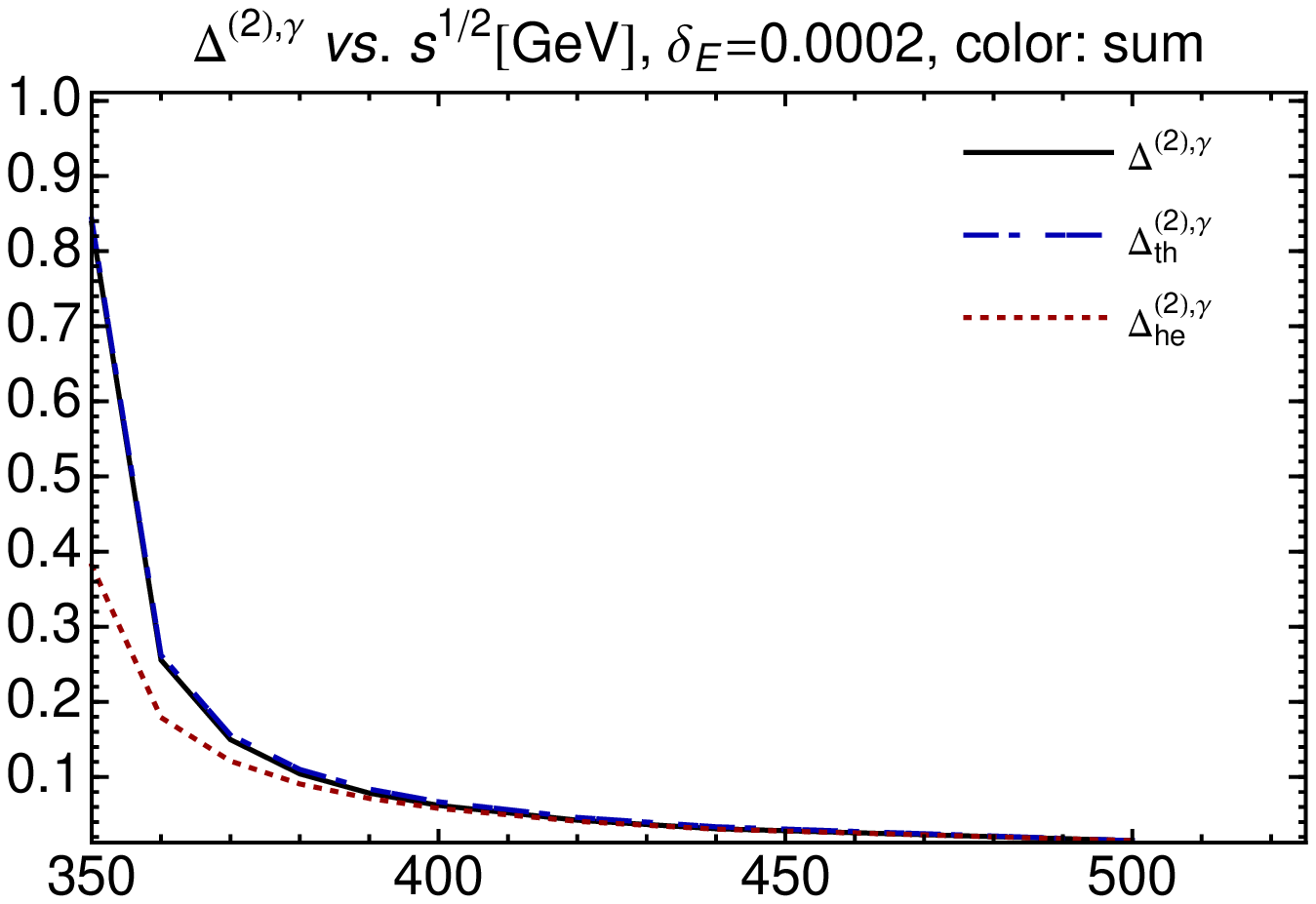}
  \includegraphics[width=0.4\textwidth]{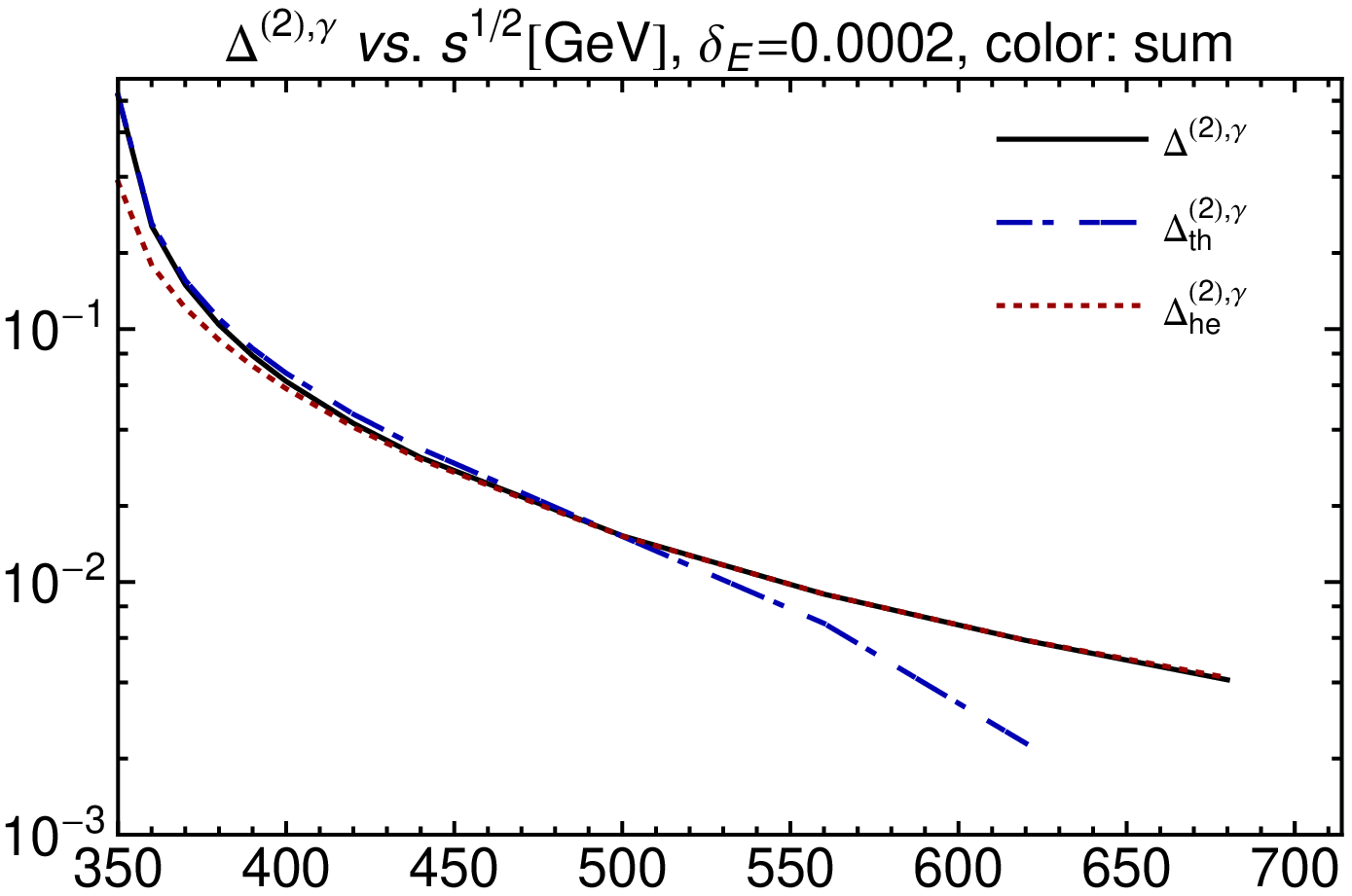}\\
  \includegraphics[width=0.42\textwidth]{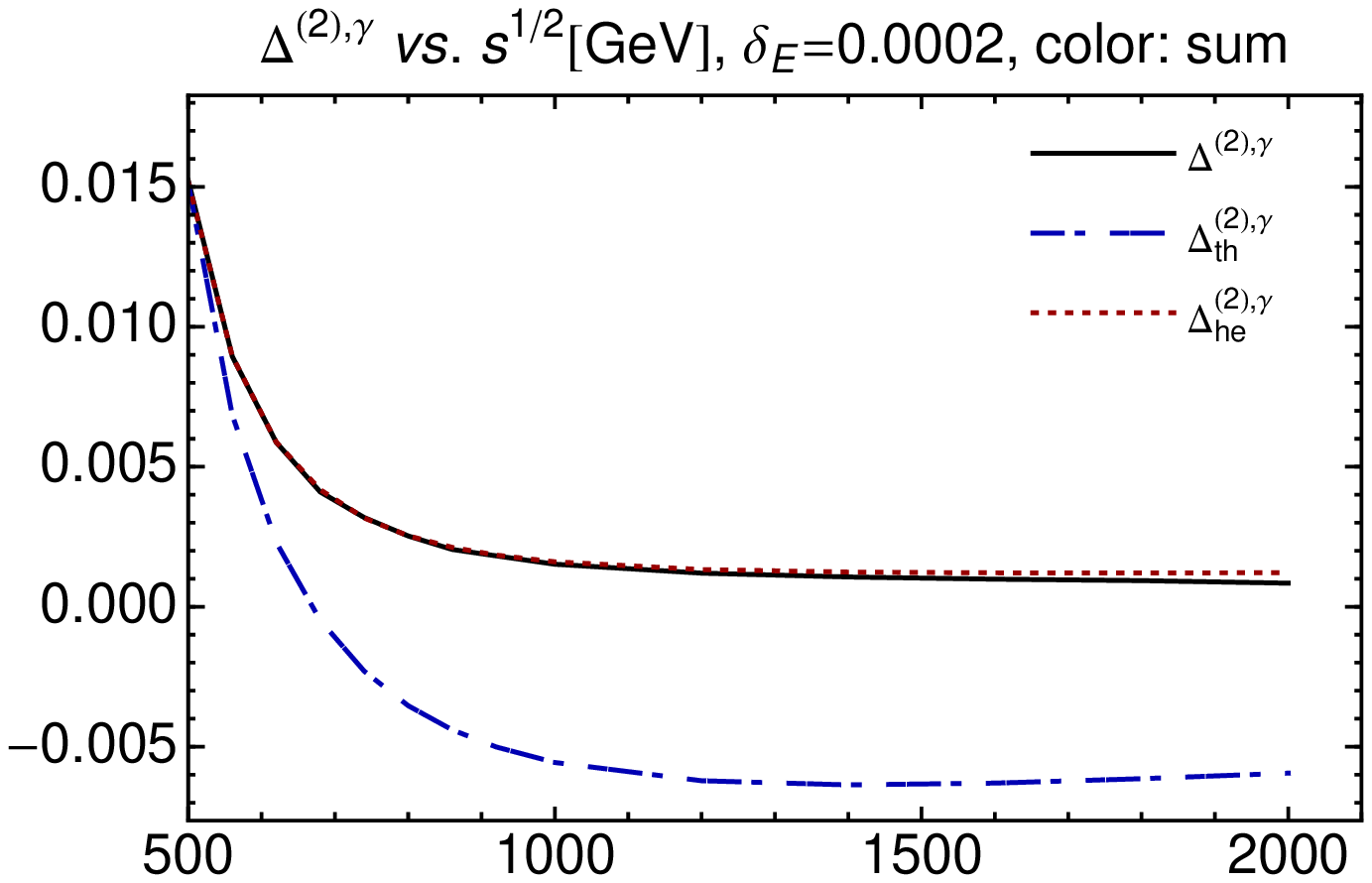}
  \end{center}
  \vspace{-1ex}
  \caption{\label{fig:scan1}
  Comparison of $\Delta^{(2),\gamma}$ with the threshold results $\Delta^{(2),\gamma}_{th}$
  and high-energy expansion results $\Delta^{(2),\gamma}_{he}$ in the
  threshold, transition, and high-energy region.}
\end{figure}

\begin{figure}[h]
  \begin{center}
  \includegraphics[width=0.4\textwidth]{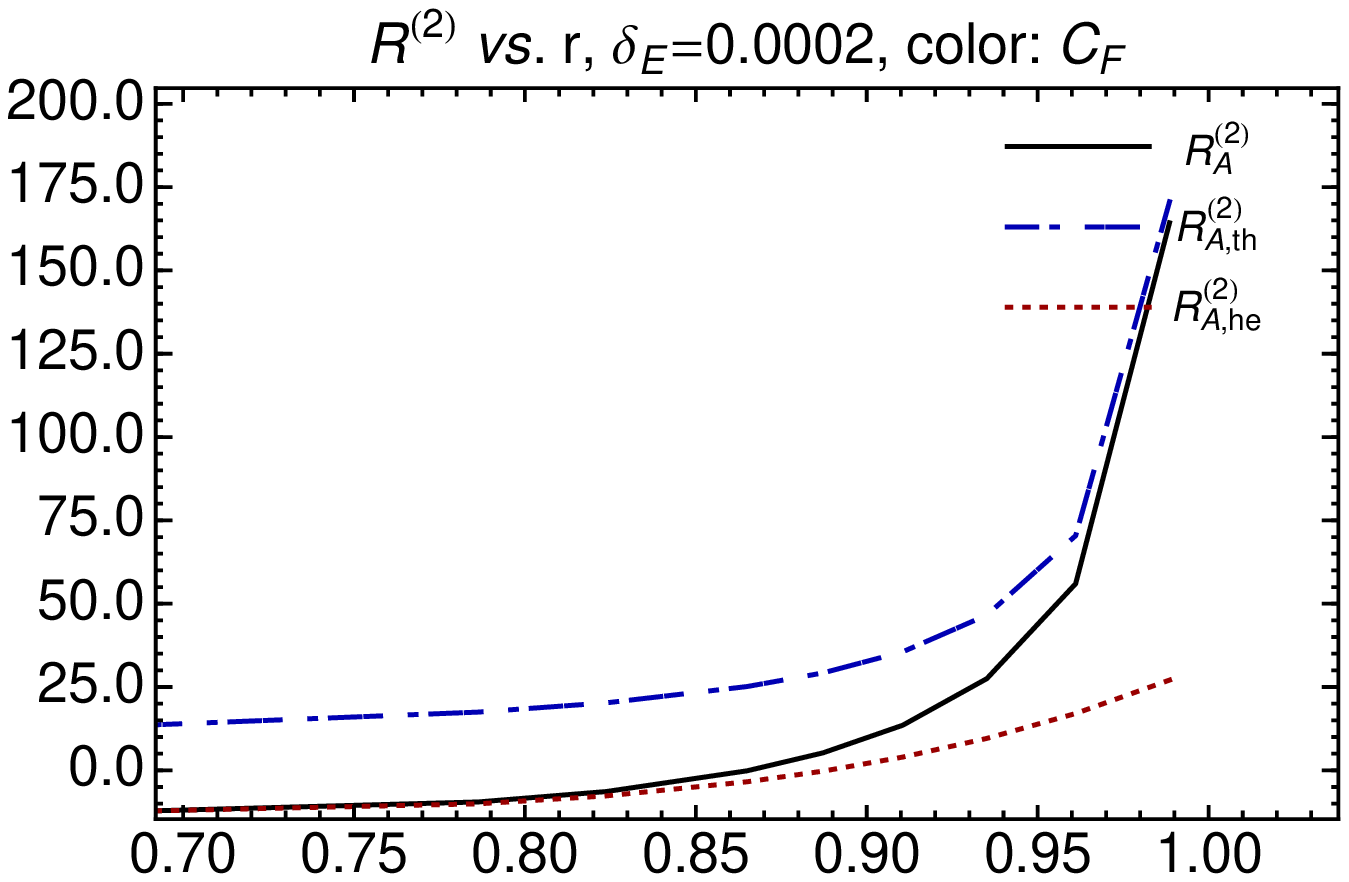}\hspace{0.5in}
  \includegraphics[width=0.4\textwidth]{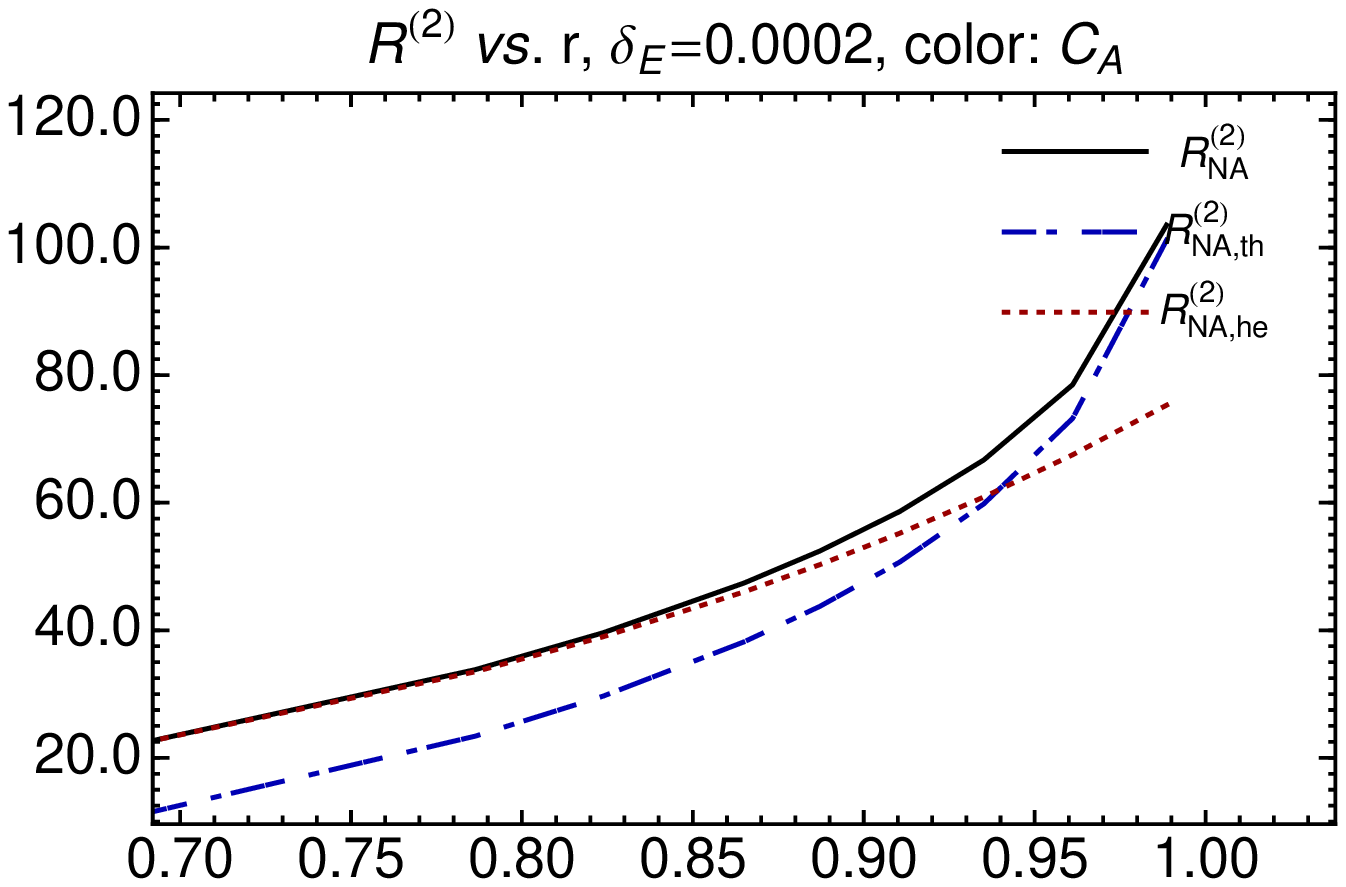}\\
  \includegraphics[width=0.4\textwidth]{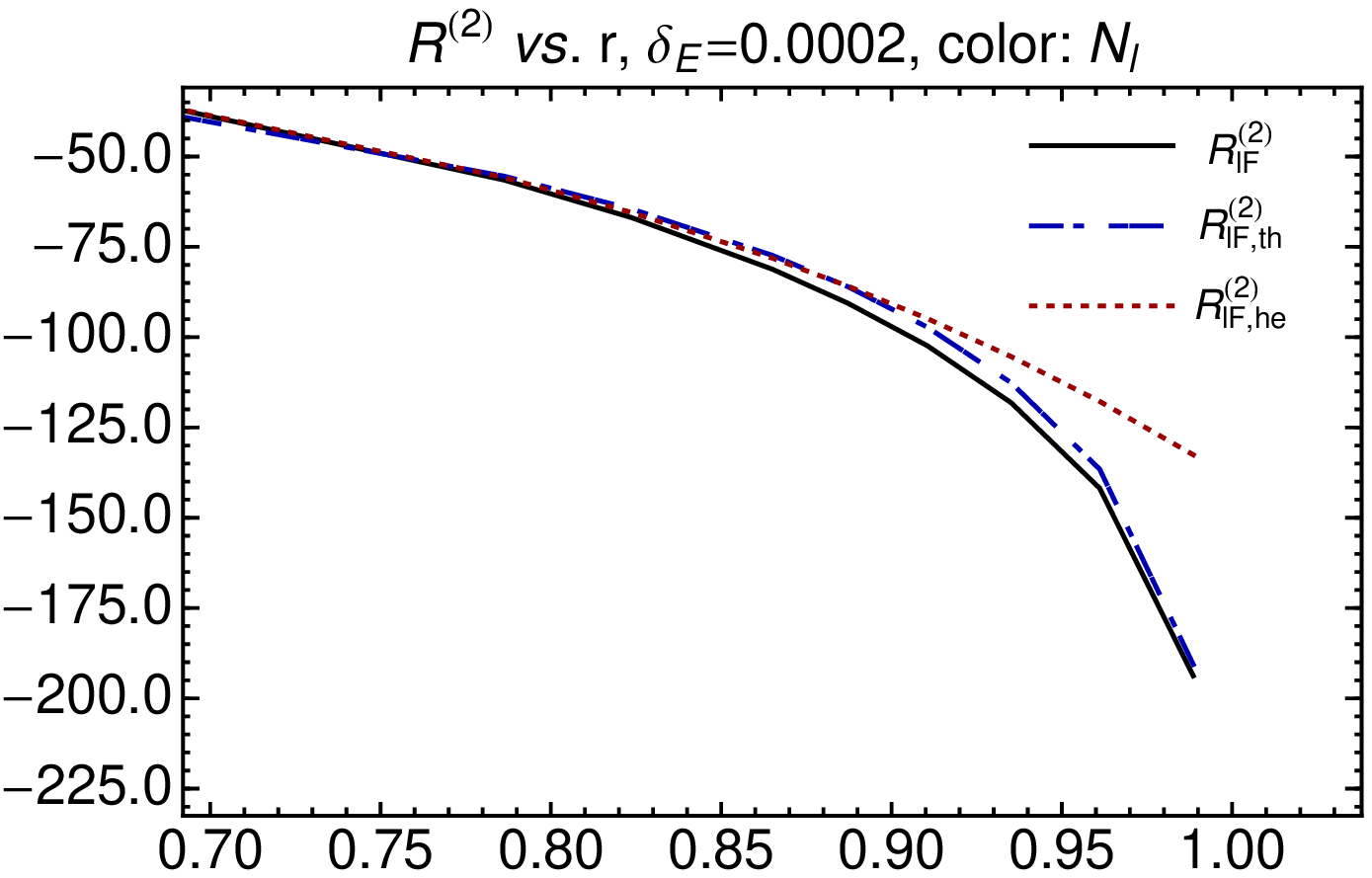}\hspace{0.5in}
  \includegraphics[width=0.4\textwidth]{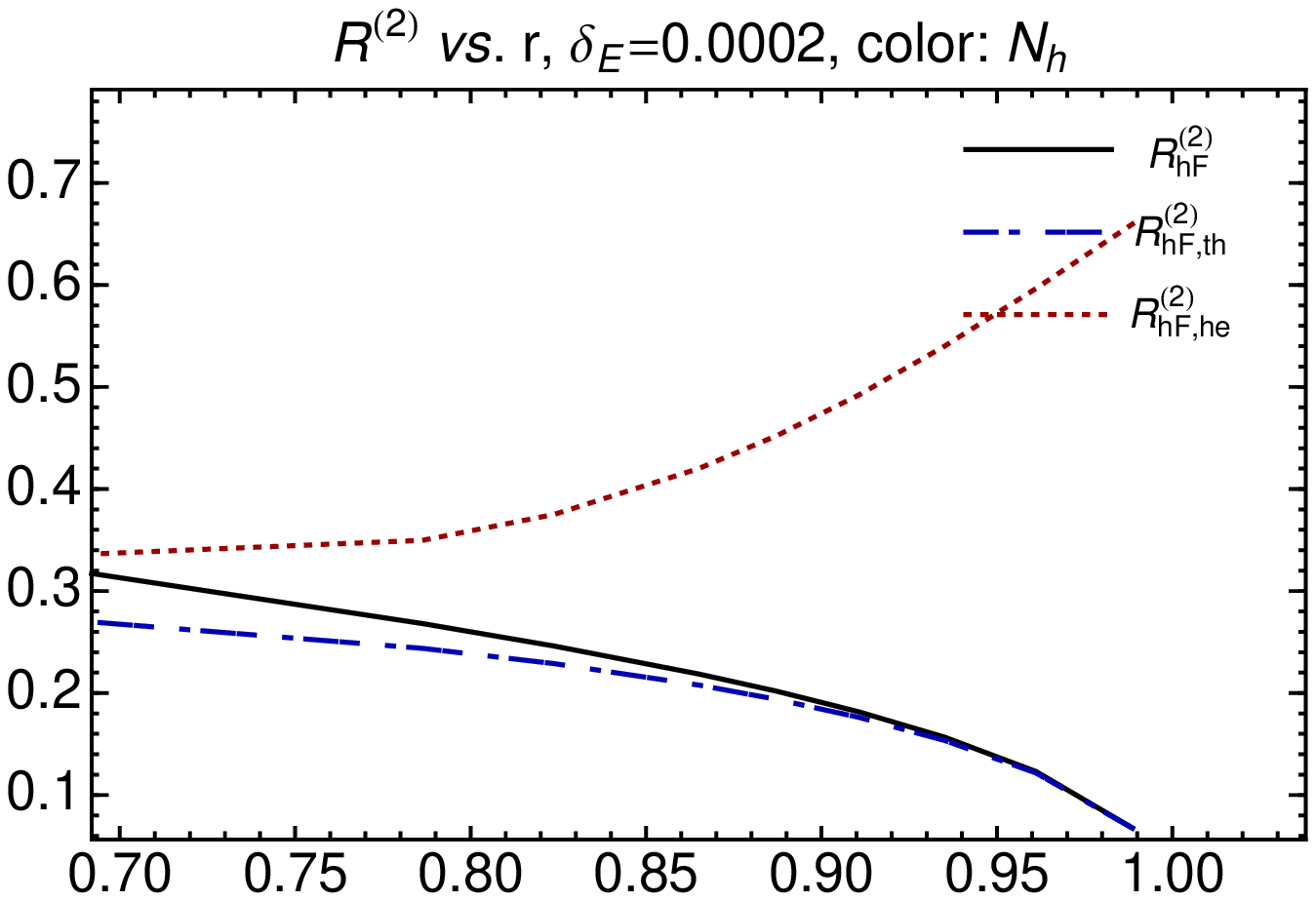}
  \end{center}
  \vspace{-1ex}
  \caption{\label{fig:scan2}
  Comparison of different color contributions of $R^{(2)}$ with the threshold results
  $R^{(2)}_{th}$ and high-energy expansion results $R^{(2)}_{he}$ in the
  threshold region.}
\end{figure}

\begin{figure}[h]
  \begin{center}
  \includegraphics[width=0.4\textwidth]{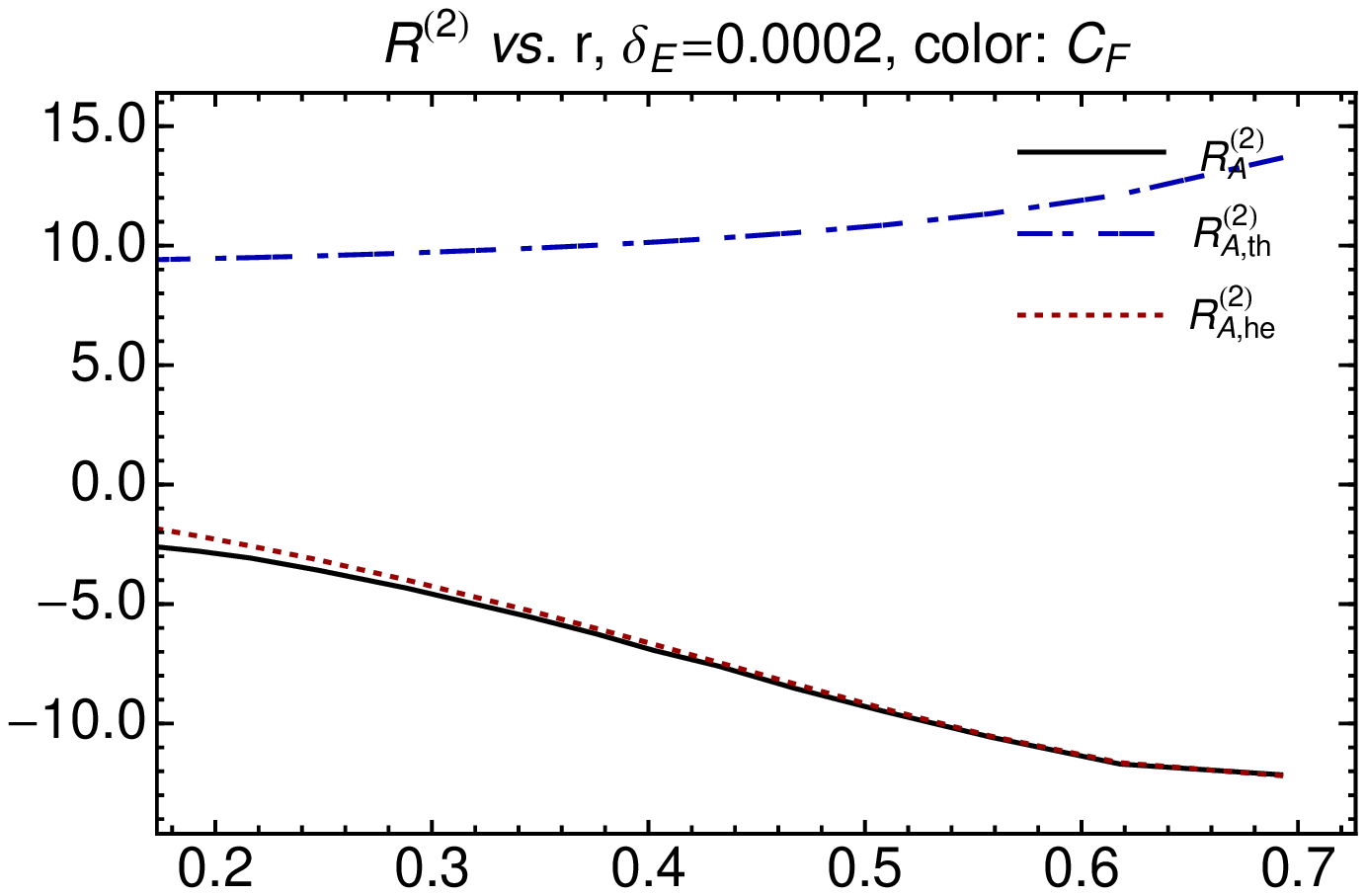}\hspace{0.5in}
  \includegraphics[width=0.4\textwidth]{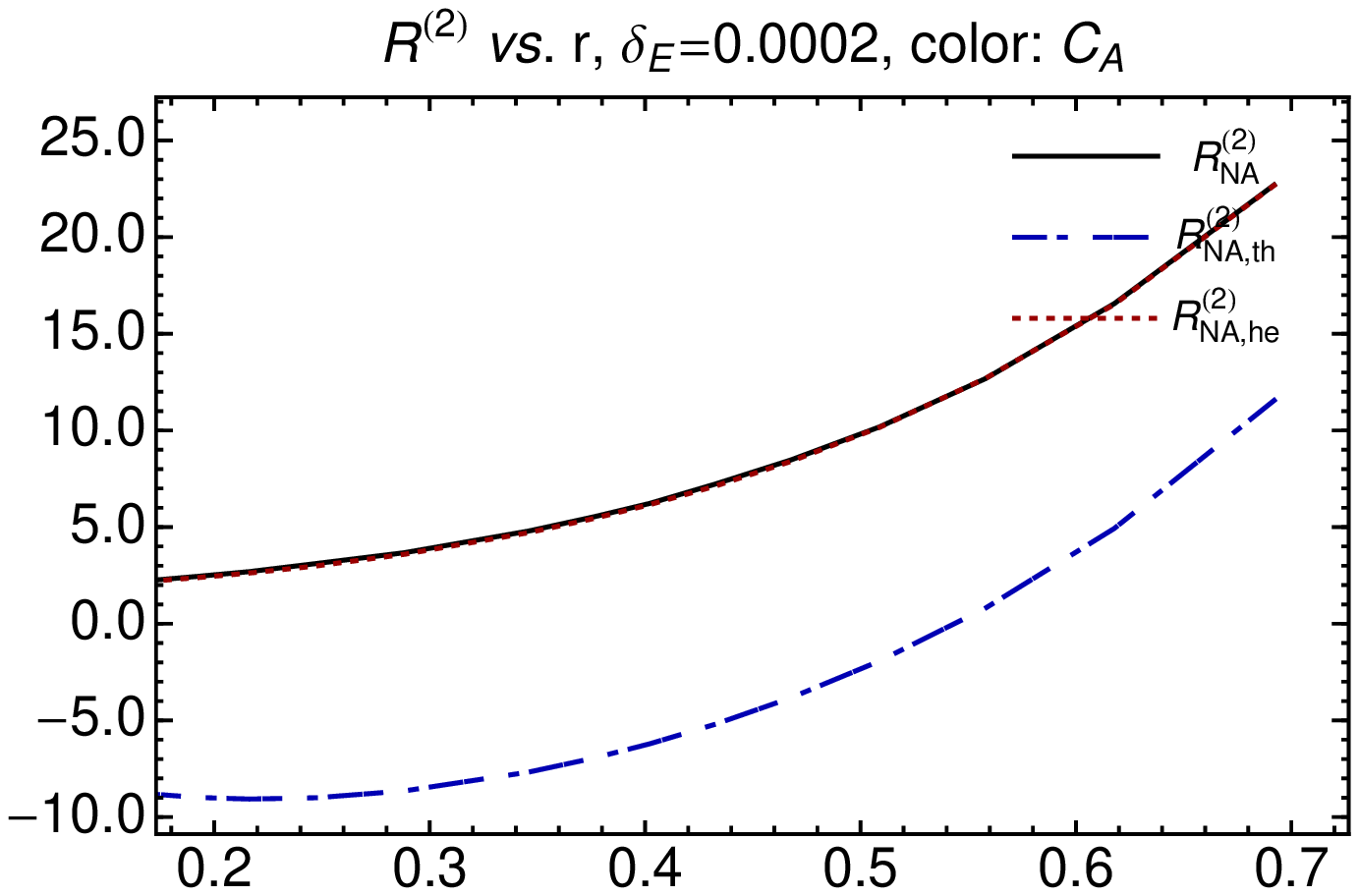}\\
  \includegraphics[width=0.4\textwidth]{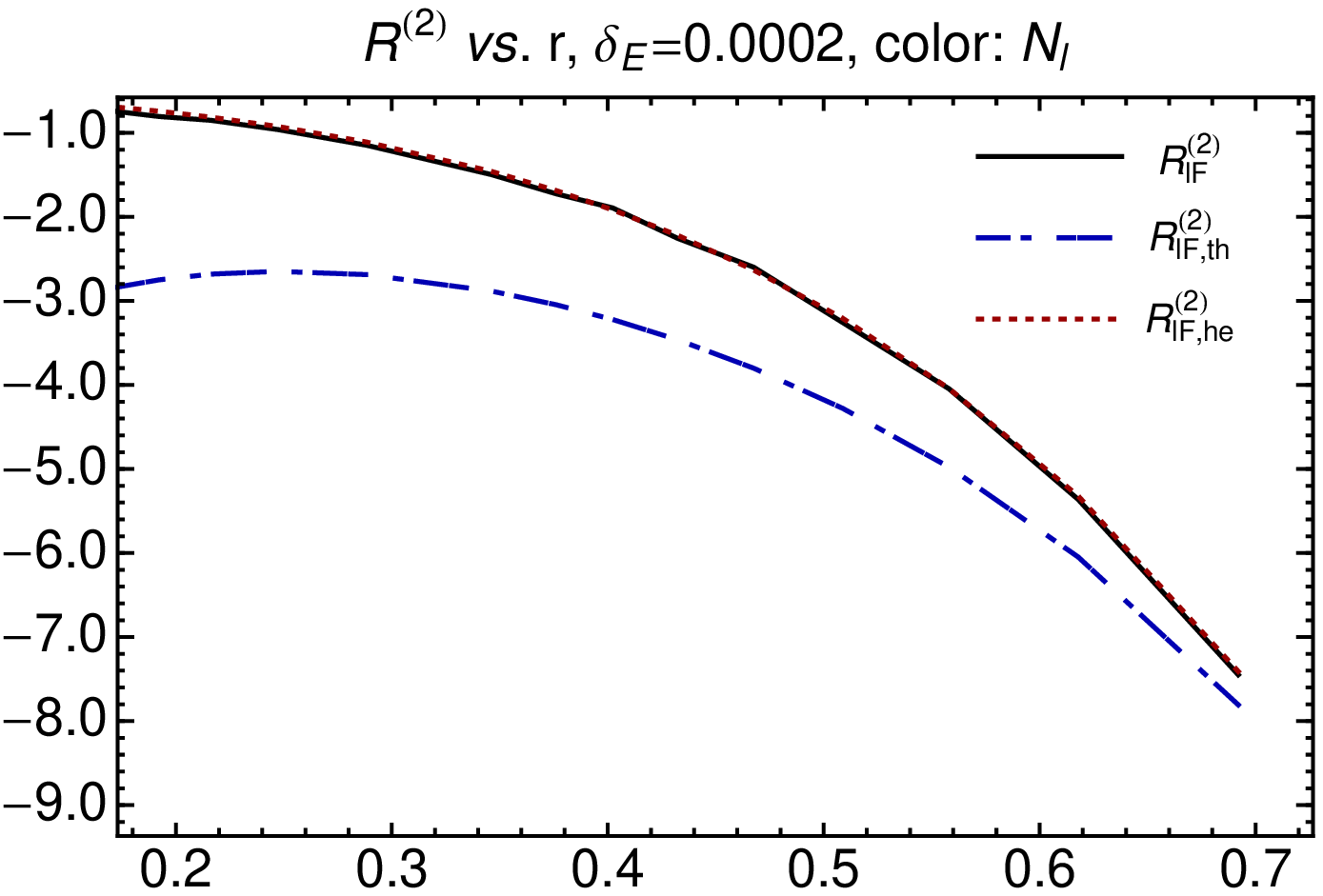}\hspace{0.5in}
  \includegraphics[width=0.4\textwidth]{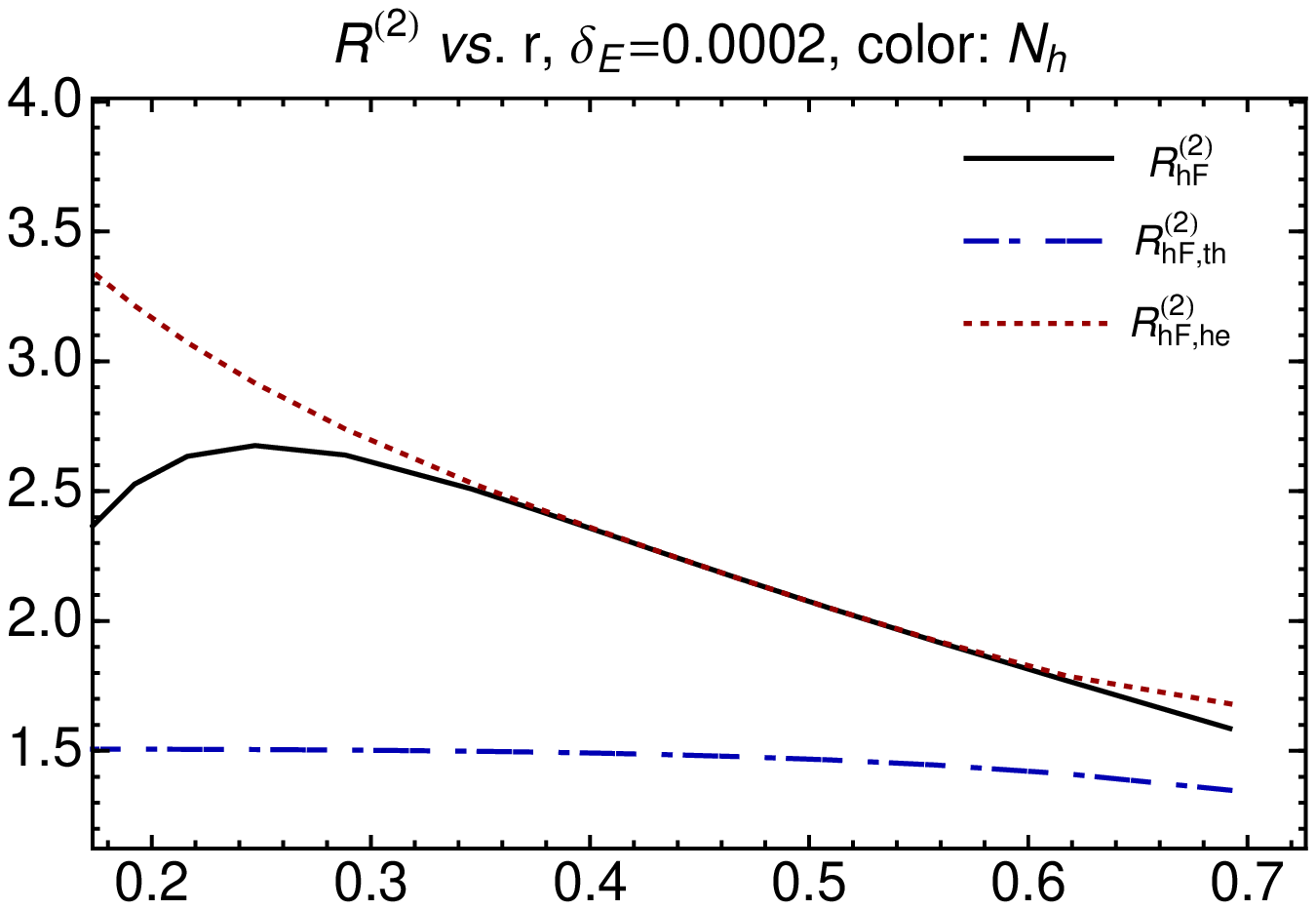}
  \end{center}
  \vspace{-1ex}
  \caption{\label{fig:scan3}
  Comparison of different color contributions of $R^{(2)}$ with the threshold results
  $R^{(2)}_{th}$ and high-energy expansion results $R^{(2)}_{he}$ in the
  high-energy region.}
\end{figure}

\begin{figure}[h]
  \begin{center}
  \includegraphics[width=0.4\textwidth]{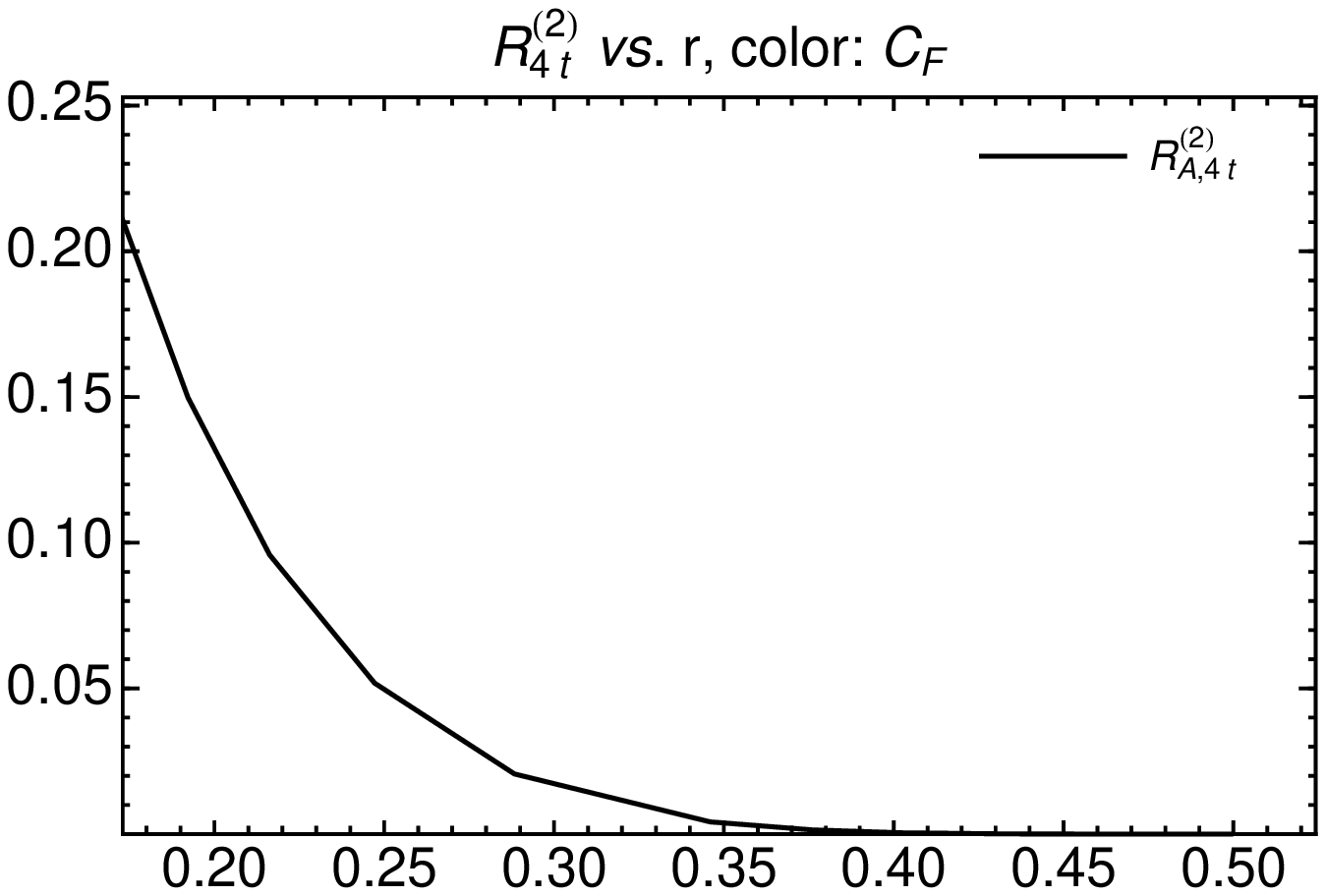}
  \includegraphics[width=0.4\textwidth]{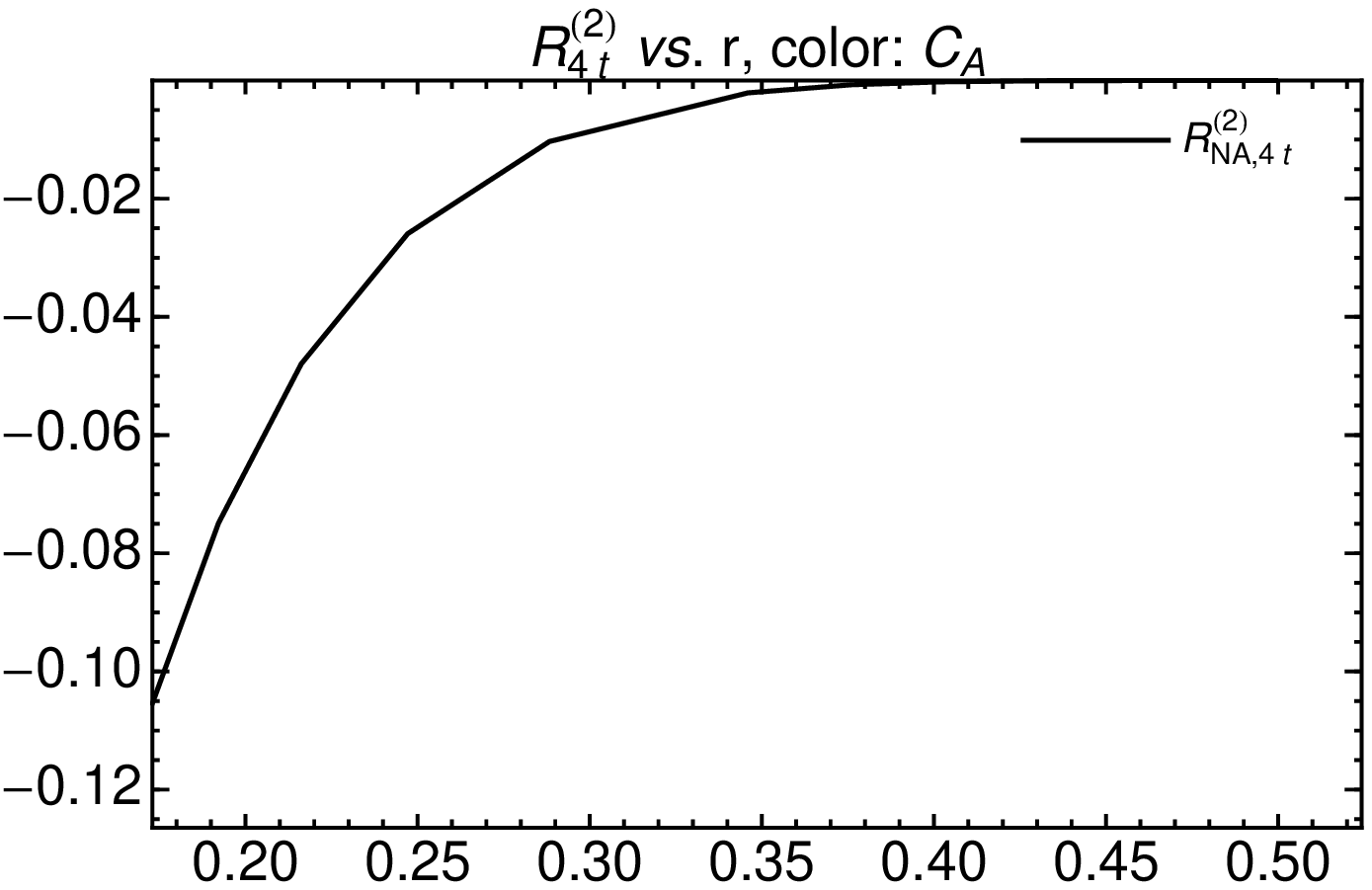} \\
  \includegraphics[width=0.4\textwidth]{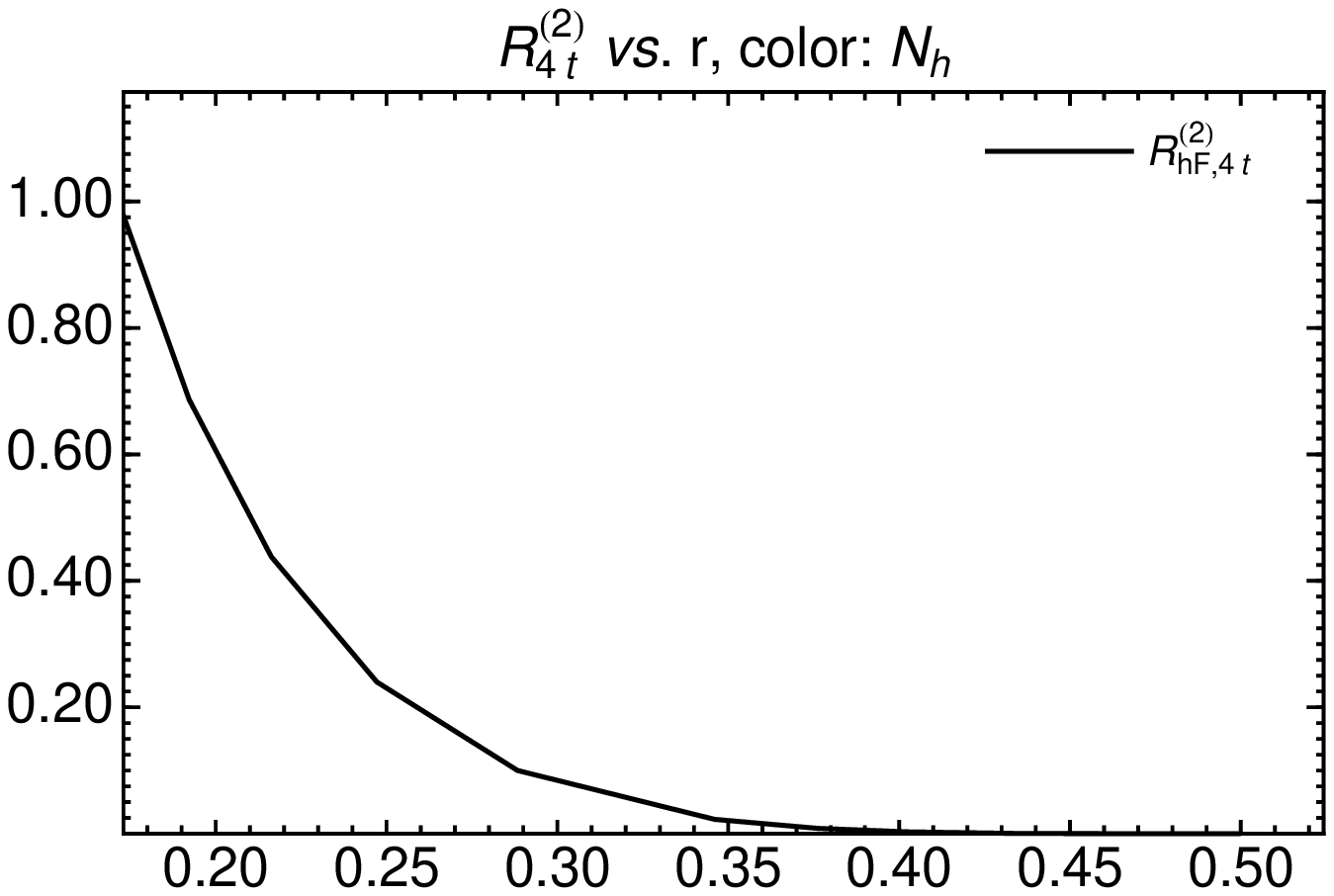}
  \end{center}
  \vspace{-1ex}
  \caption{\label{fig:scan4}
  Contributions to $R^{(2)}$ with different color structure from double-real corrections   of four top quark final states.}
\end{figure}

Fig.~\ref{fig:scan1} shows detailed comparison of our numerical results
with the threshold~\cite{hep-ph/9712222,hep-ph/9712302} and
high-energy expansion results~\cite{hep-ph/9710413,hep-ph/9704222} in the
threshold, transition, and high-energy region for a fixed $\delta_E=2\times10^{-4}$.
It can be seen that our full results works well in the entire energy region,
i.e., approaching the threshold results for lower energies and the high-energy
expansions on another end, while the other twos are not. However, one may notice
the differences between the high-energy expansion results and ours for $\sqrt s>1200$ GeV.
Though the differences are only at a level of a few times $10^{-4}$.
The comparison are also shown in terms of $R^{(2)}$ in Figs.~\ref{fig:scan2}-\ref{fig:scan3}
as functions of $r$ for different color contributions. From Fig.~\ref{fig:scan3} we
can see more clearly the differences in regions of $r < 0.3$, especially for the Abelian
and heavy-fermionic contributions. Most of these differences are attributed to the
inclusion of double-real corrections with four top quark final state in~\cite{hep-ph/9710413,hep-ph/9704222}, which
are not included in our calculations by default. We calculate
those contributions to $R^{(2)}$ separately as shown in Fig.~\ref{fig:scan4}, which
are only non-negligible for $r< 0.3$. They
are positive for the Abelian and heavy-fermionic parts of $R^{(2)}$ and negative for
the non-Abelian part. These four top contributions have been checked against~\cite{hep-ph/9705295} and
found in very good agreement. Another reason is because we choose a finite value of
$\delta_E=2\times 10^{-4}$, for which the $f_1$ and $f_2$ terms add up to about
$10^{-4}$ for Abelian and non-Abelian parts with $\sqrt s=2000$ GeV .

\begin{figure}[h]
  \begin{center}
  \includegraphics[width=0.4\textwidth]{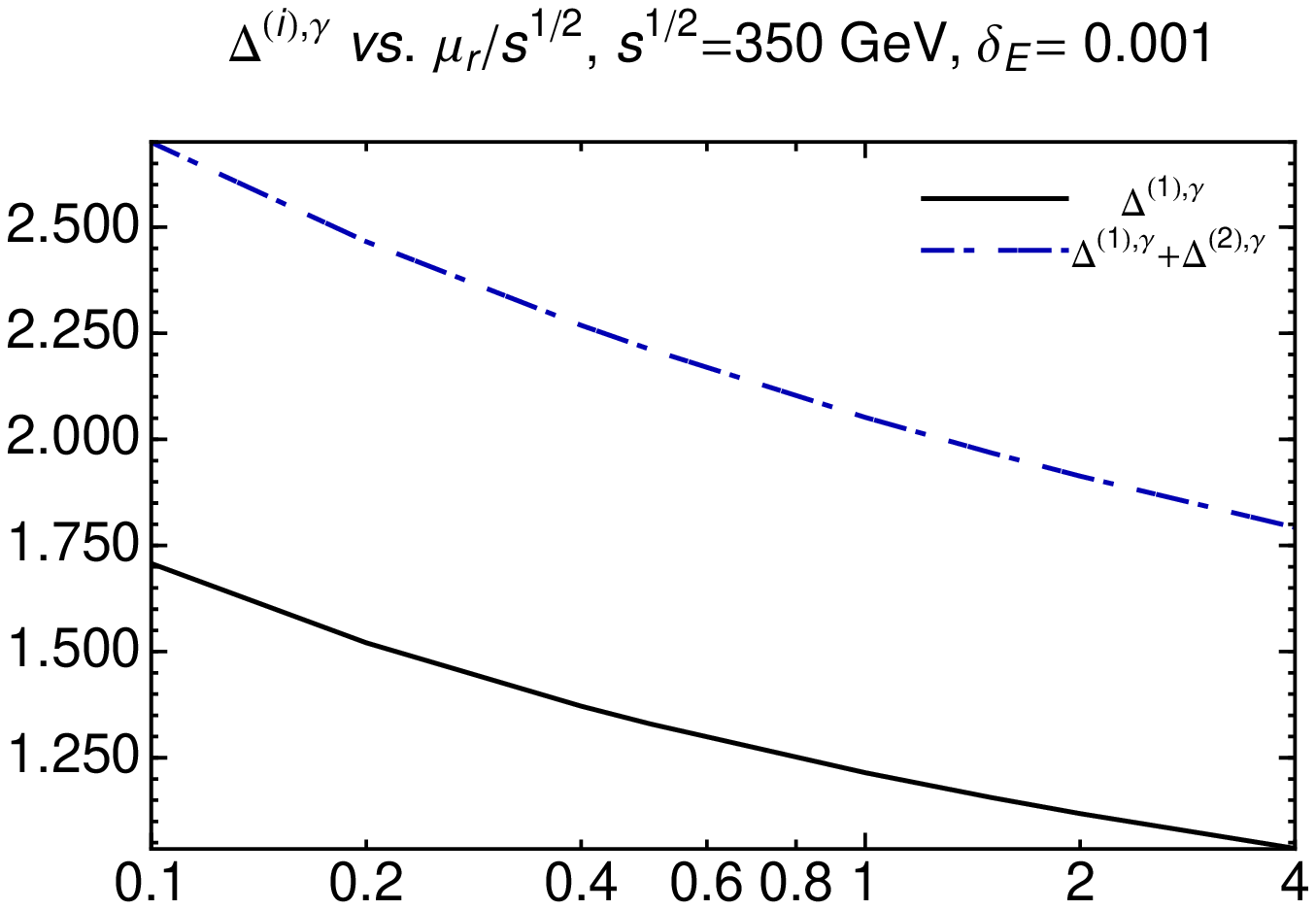}
  \includegraphics[width=0.4\textwidth]{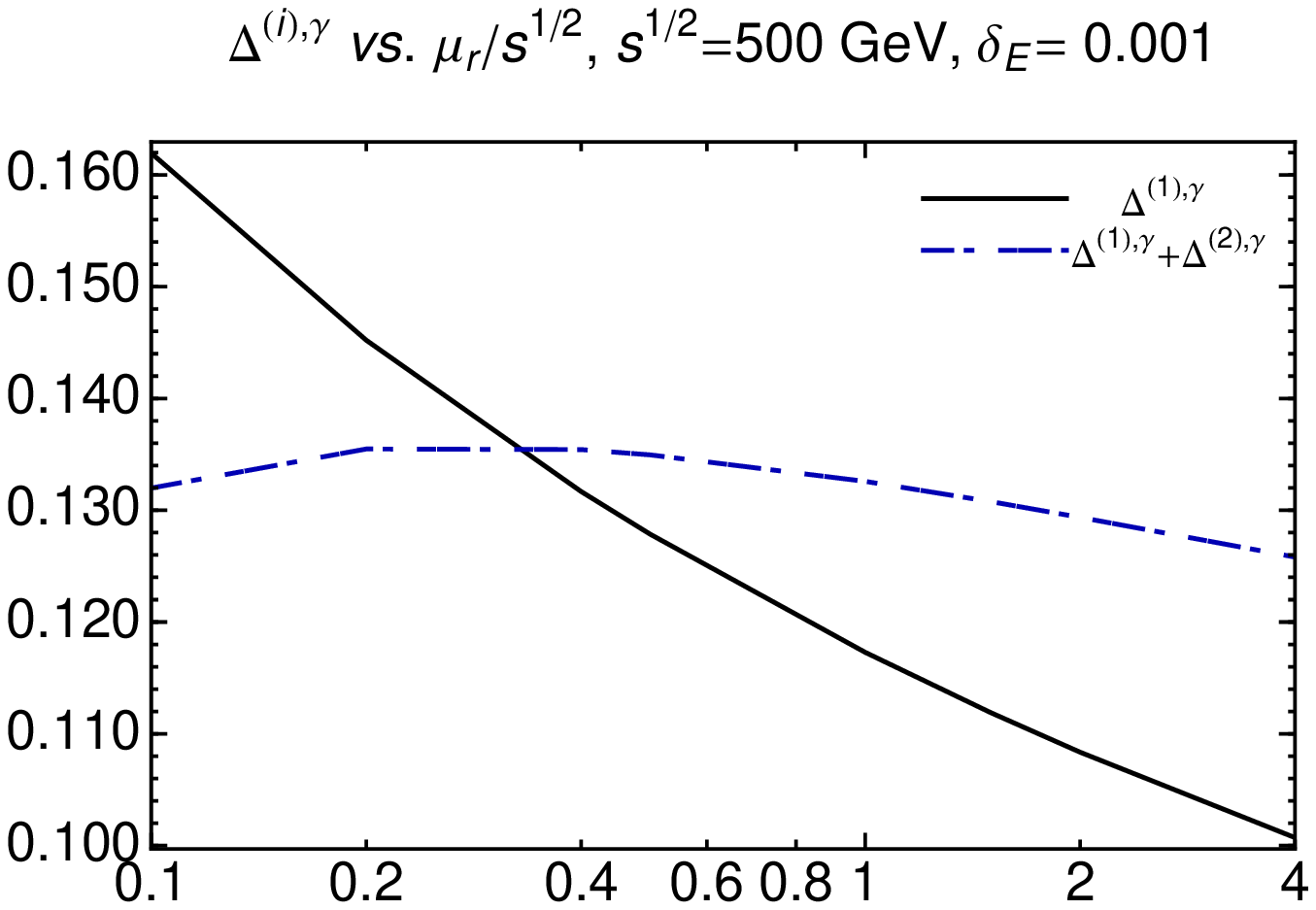} \\
  \includegraphics[width=0.4\textwidth]{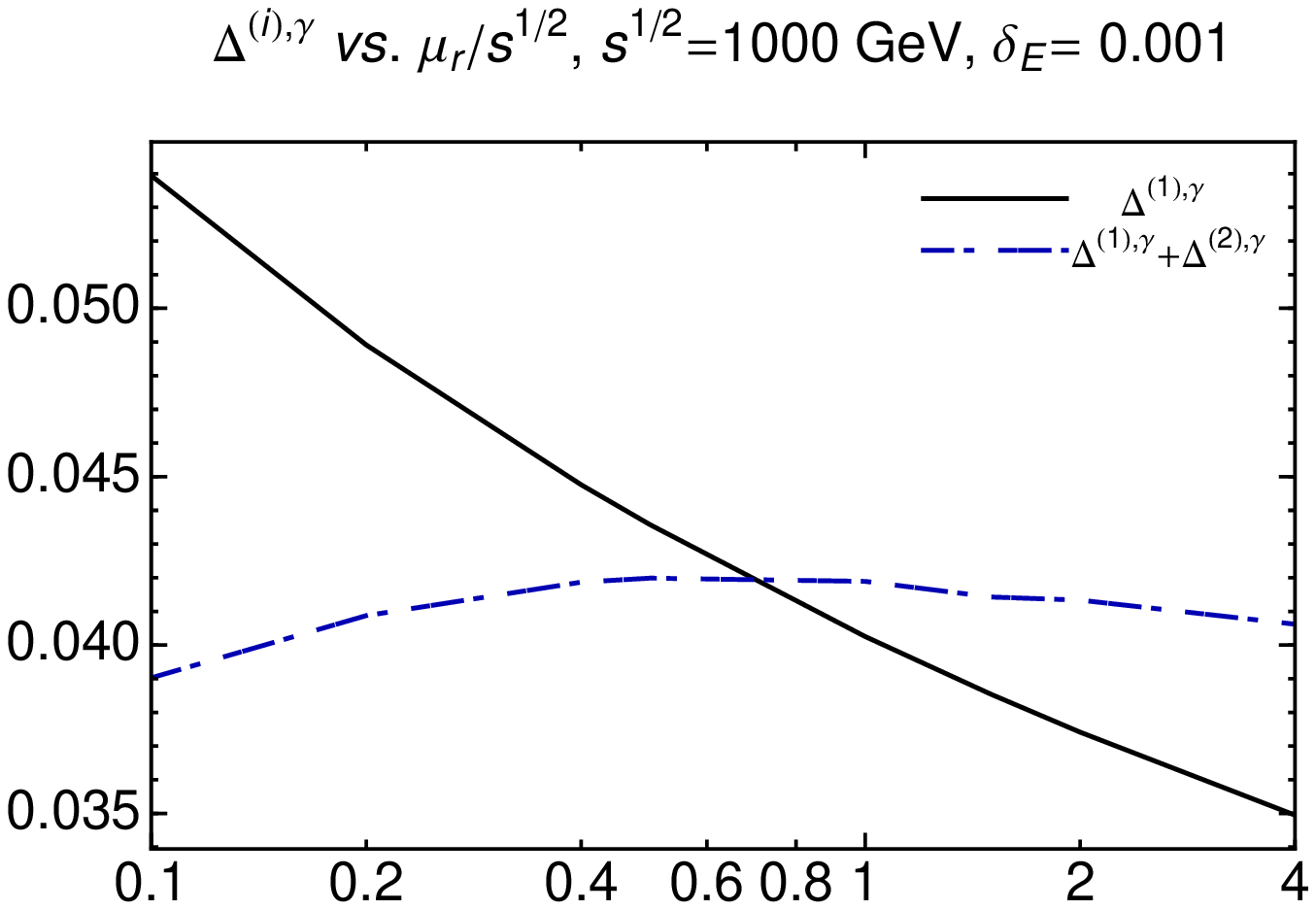}
  \end{center}
  \vspace{-1ex}
  \caption{\label{fig:scale1}
  Scale dependence of $\Delta^{(1),\gamma}$ and $\Delta^{(1),\gamma}+\Delta^{(2),\gamma}$
  for different collision energies.}
\end{figure}

We further show reduction of the scale variations by including the $\ordb$ corrections
in Fig.~\ref{fig:scale1}. We vary the renormalization scale $\mu_r$ around the nominal
choice $\mu_r=\sqrt s$ by a factor of 10 downward and 4 upward. The scale dependence
have been reduced significantly for $\sqrt s=$ 500 and 1000 GeV, e.g., from 6\% at
the NLO to 1\% at the NNLO for a collision energy of 500 GeV. The NNLO results still
show a large scale dependence near production threshold due to the large corrections
and require resummations for further improvements.

\subsection{Differential distributions}
\begin{figure}[h]
  \begin{center}
  \includegraphics[width=0.4\textwidth]{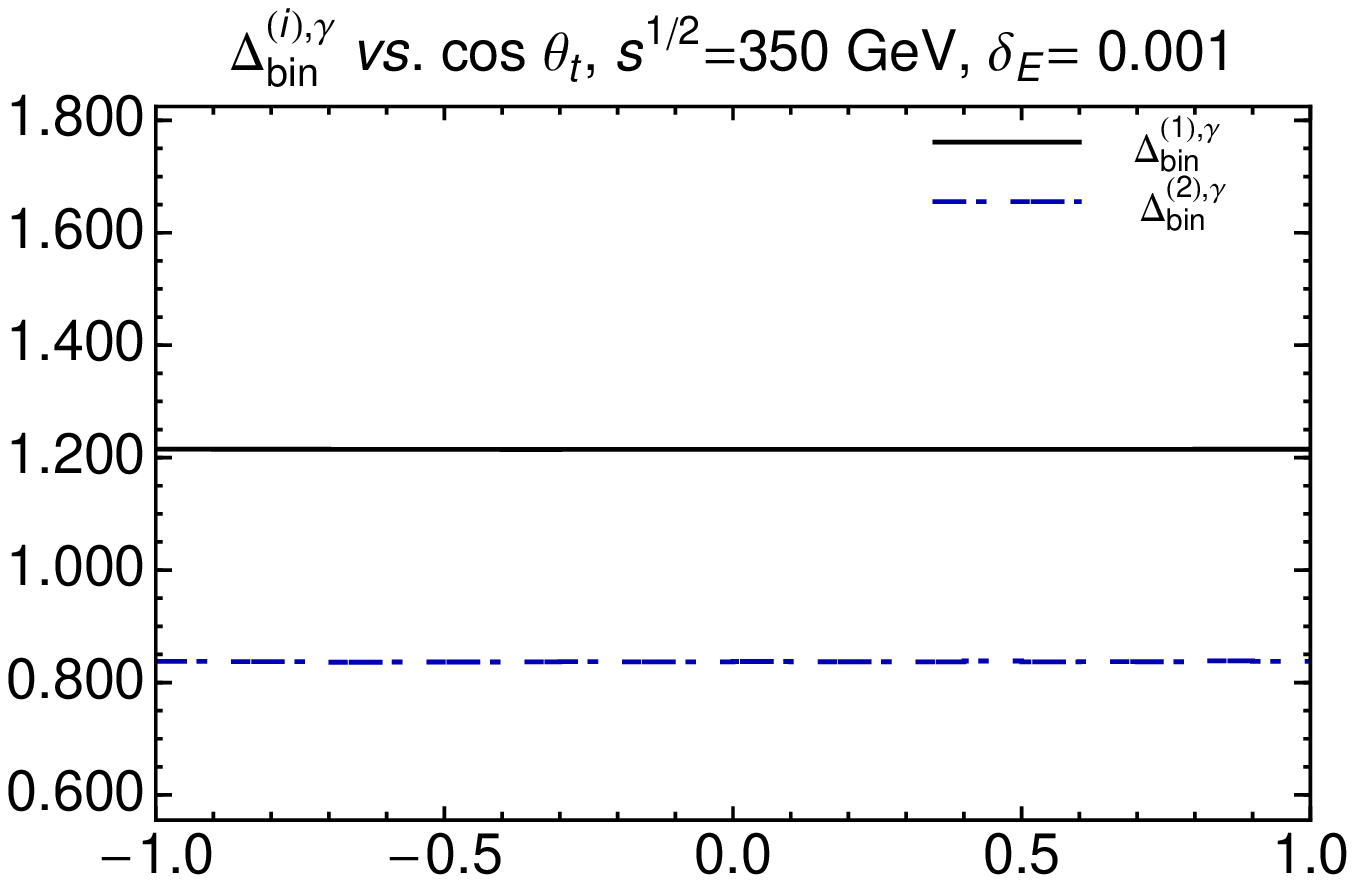}
  \includegraphics[width=0.4\textwidth]{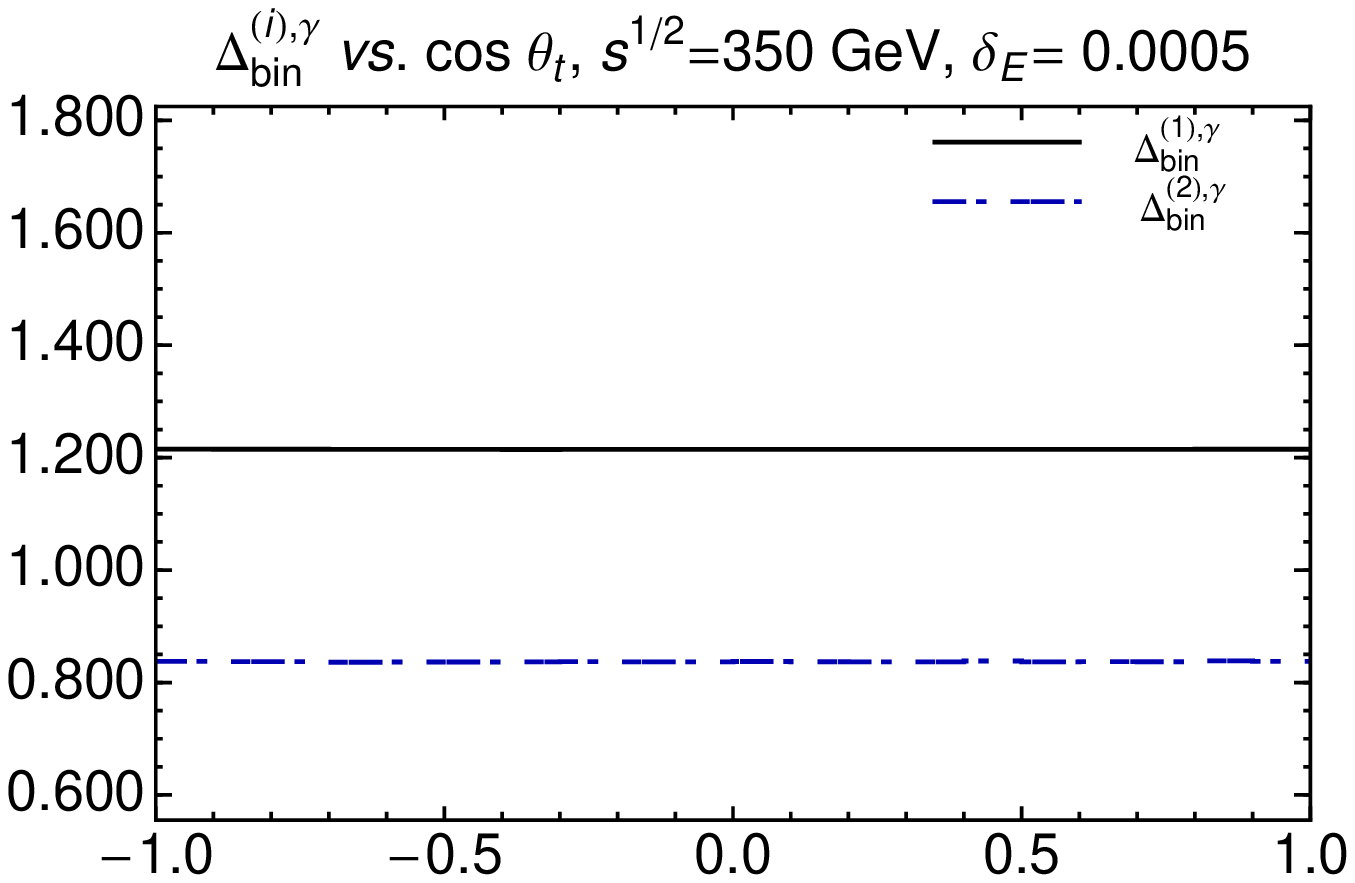} \\
\includegraphics[width=0.4\textwidth]{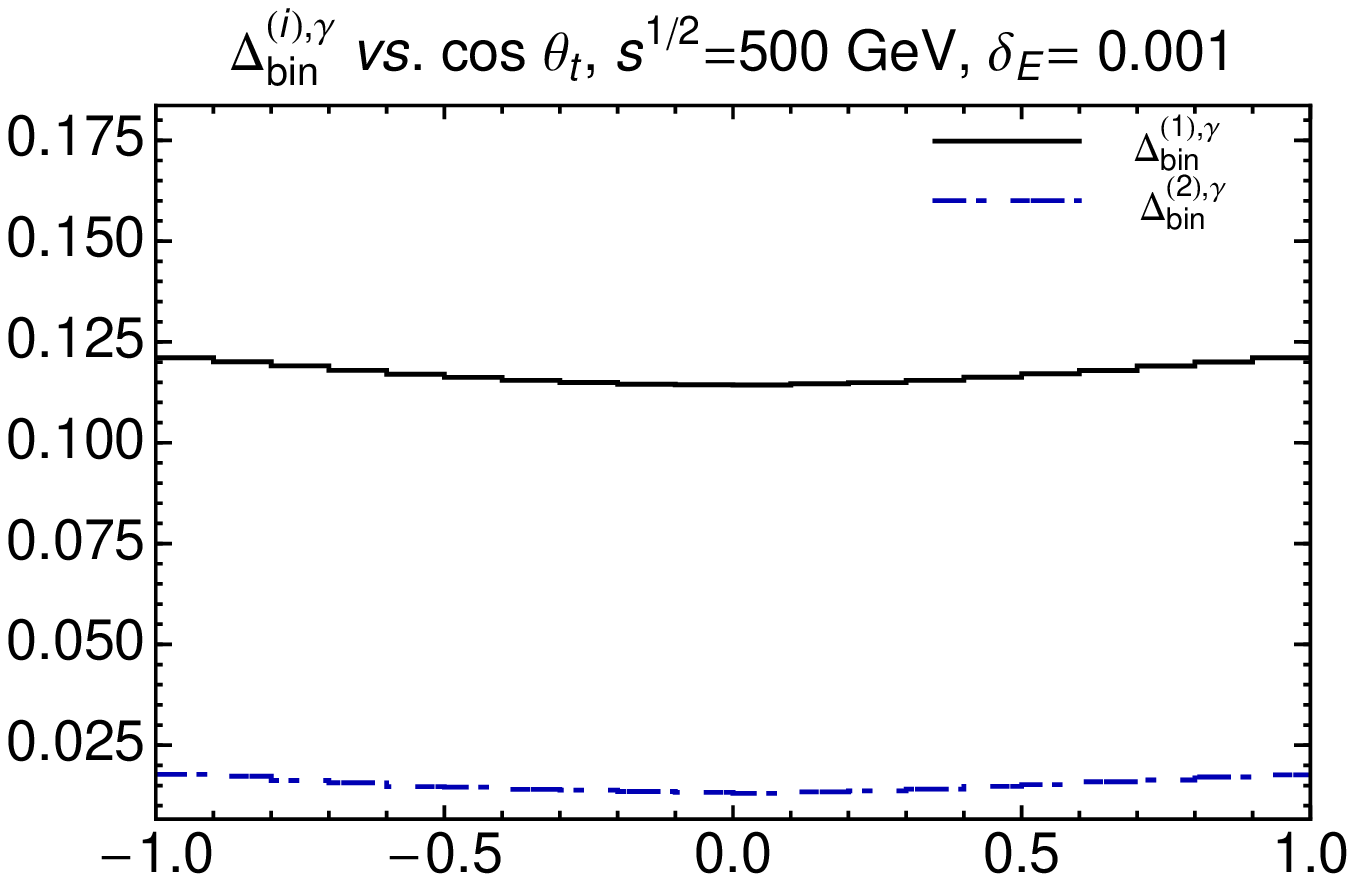}
\includegraphics[width=0.4\textwidth]{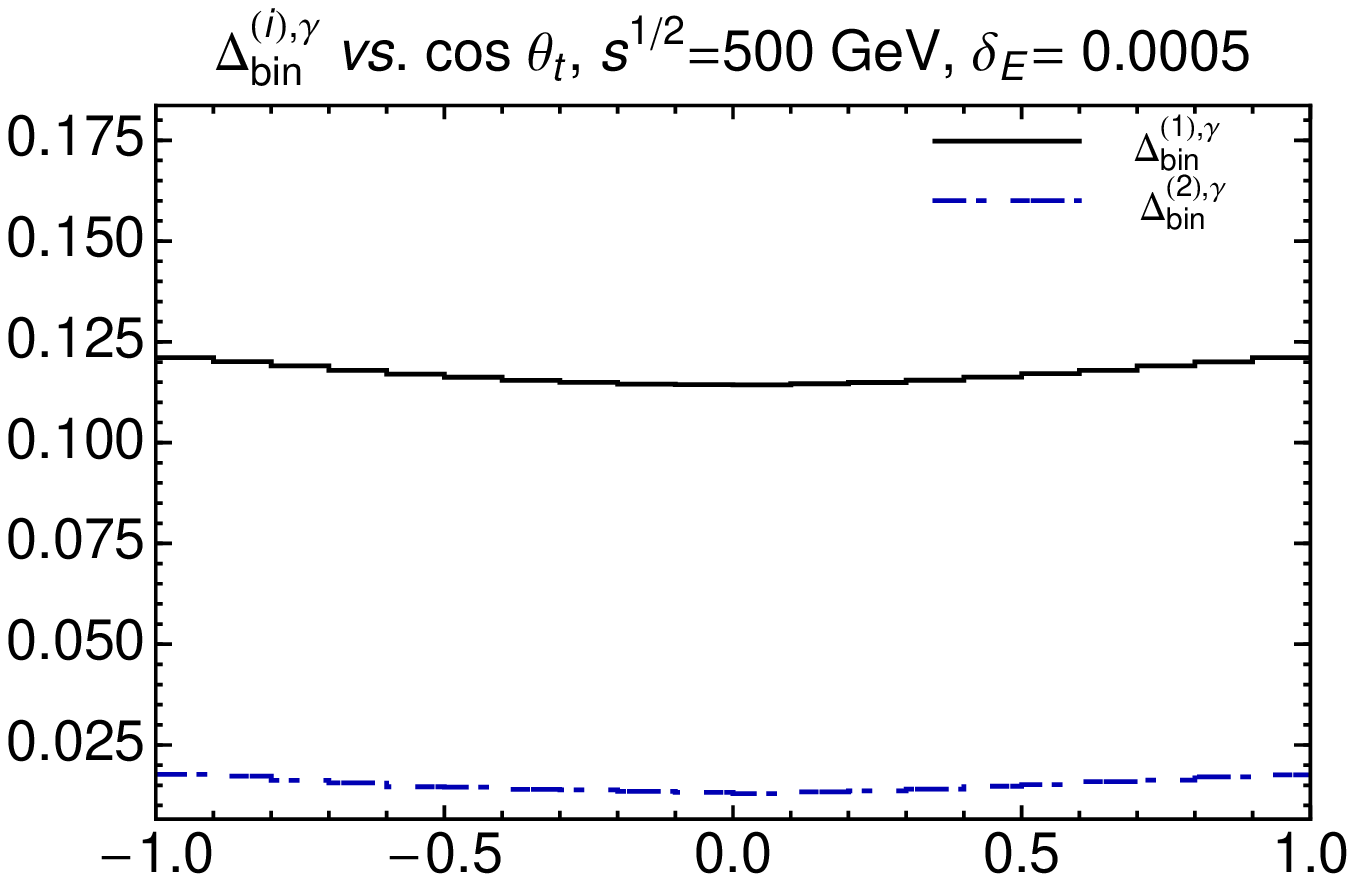}
\\
  \includegraphics[width=0.4\textwidth]{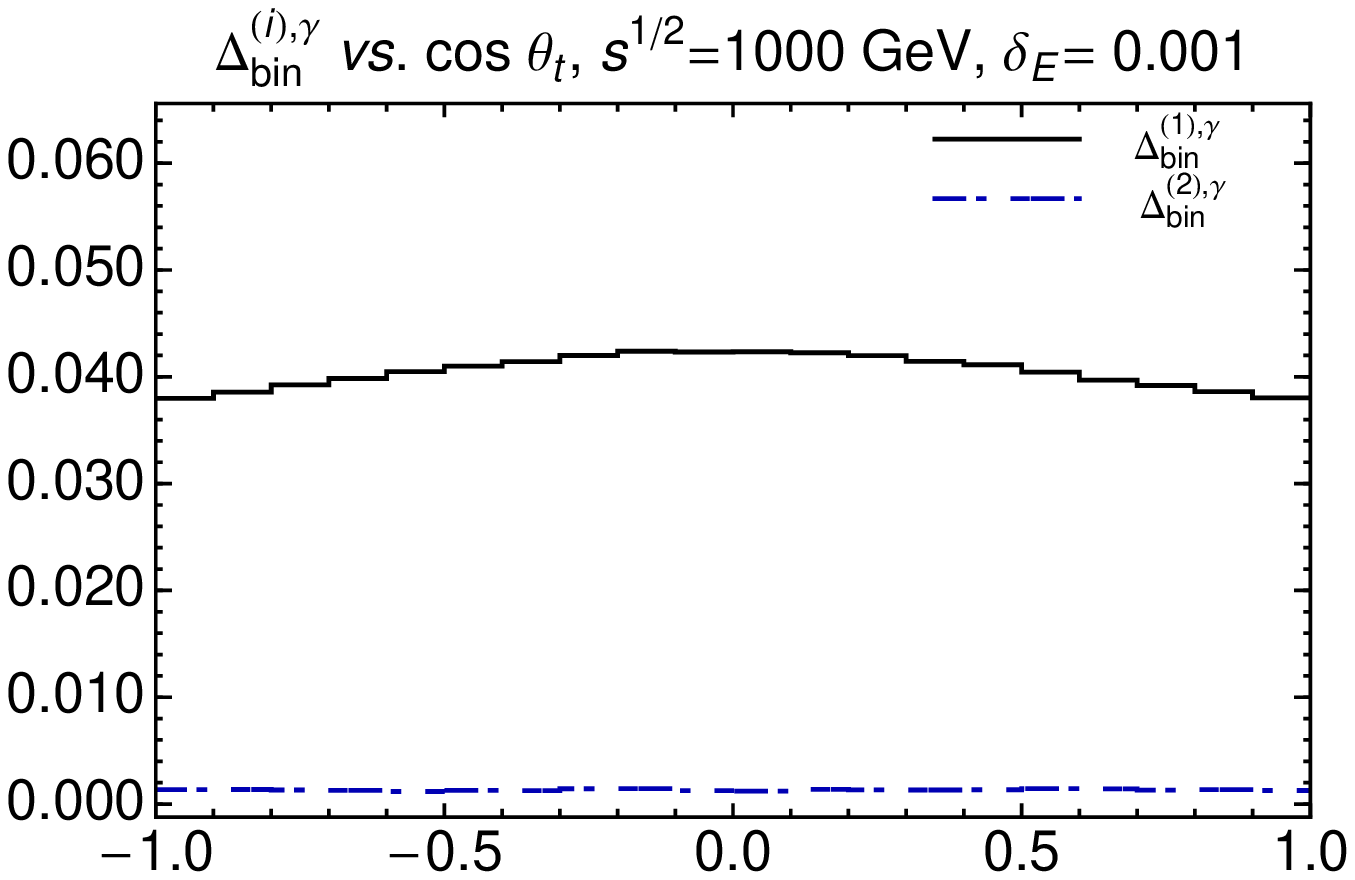}
  \includegraphics[width=0.4\textwidth]{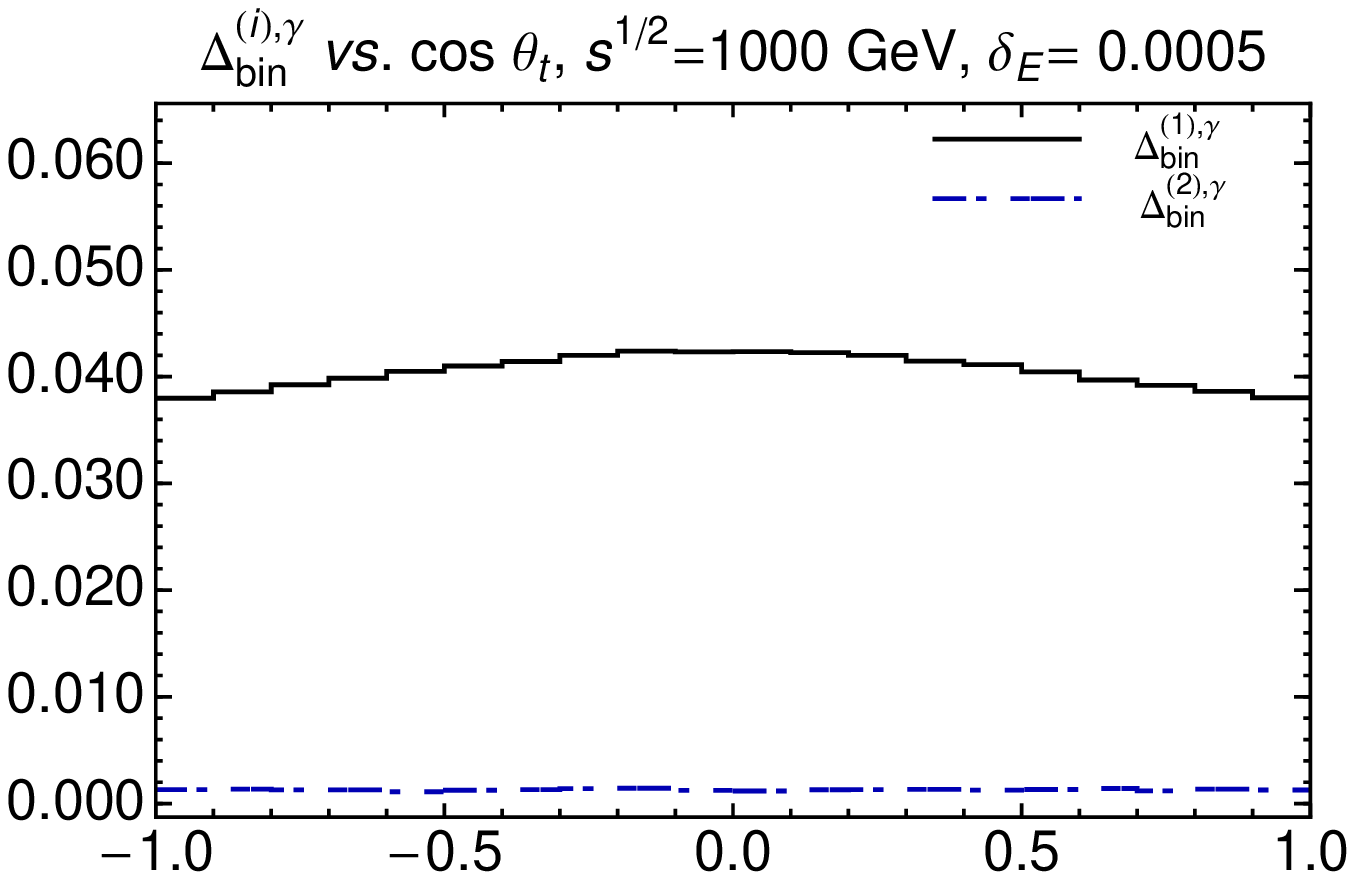}
  \end{center}
  \vspace{-1ex}
  \caption{\label{fig:dis1}
  $\mathcal{O}(\als)$ and $\mathcal{O}(\als^2)$ corrections in different $\cos\theta_t$ bins, $\Delta^{(1),\gamma}_{bin}$ and $\Delta^{(2),\gamma}_{bin}$
  for different collision energies and different $\delta_E$ choices.}
\end{figure}

\begin{figure}[h]
  \begin{center}
  \includegraphics[width=0.4\textwidth]{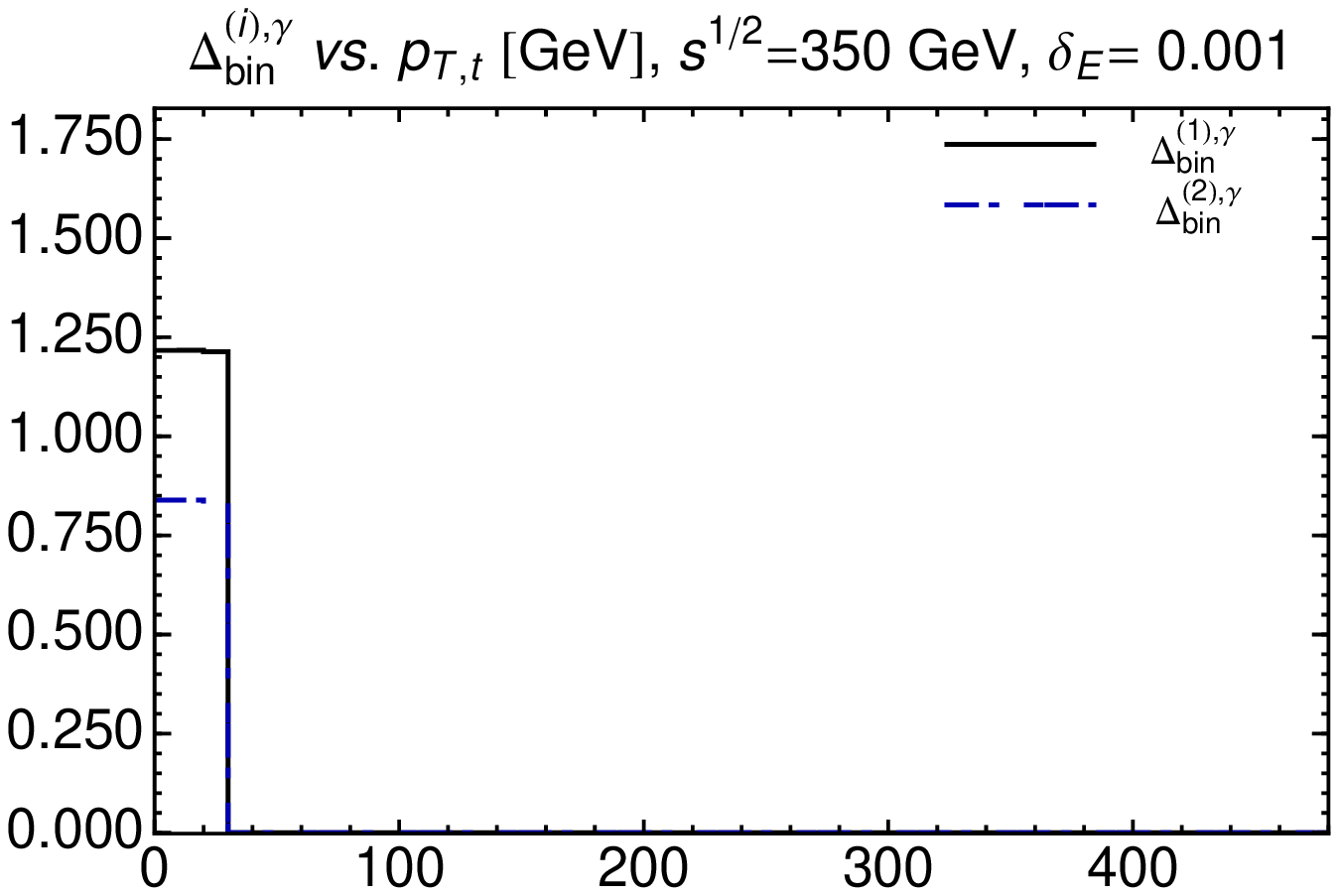}
  \includegraphics[width=0.4\textwidth]{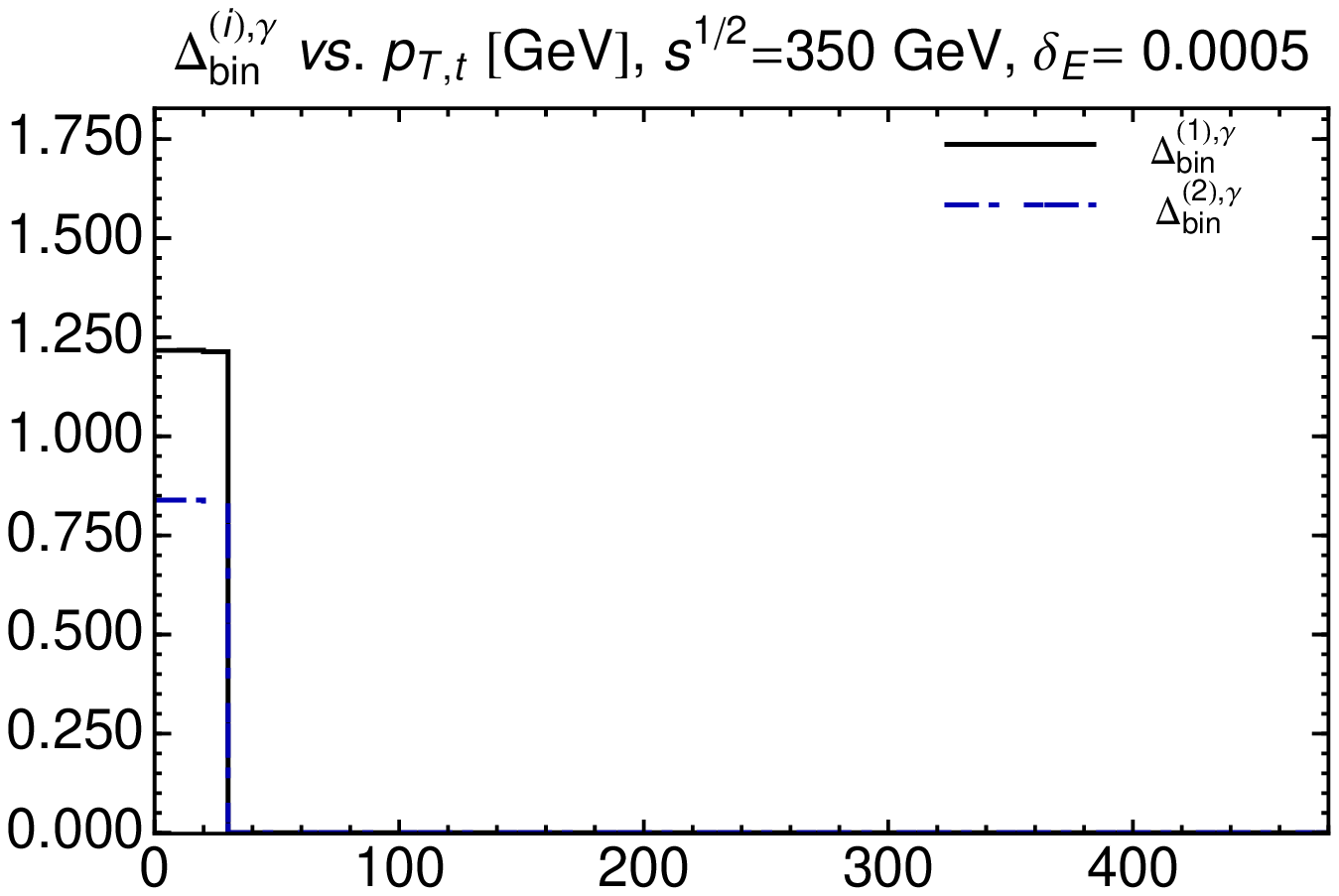}\\  
\includegraphics[width=0.4\textwidth]{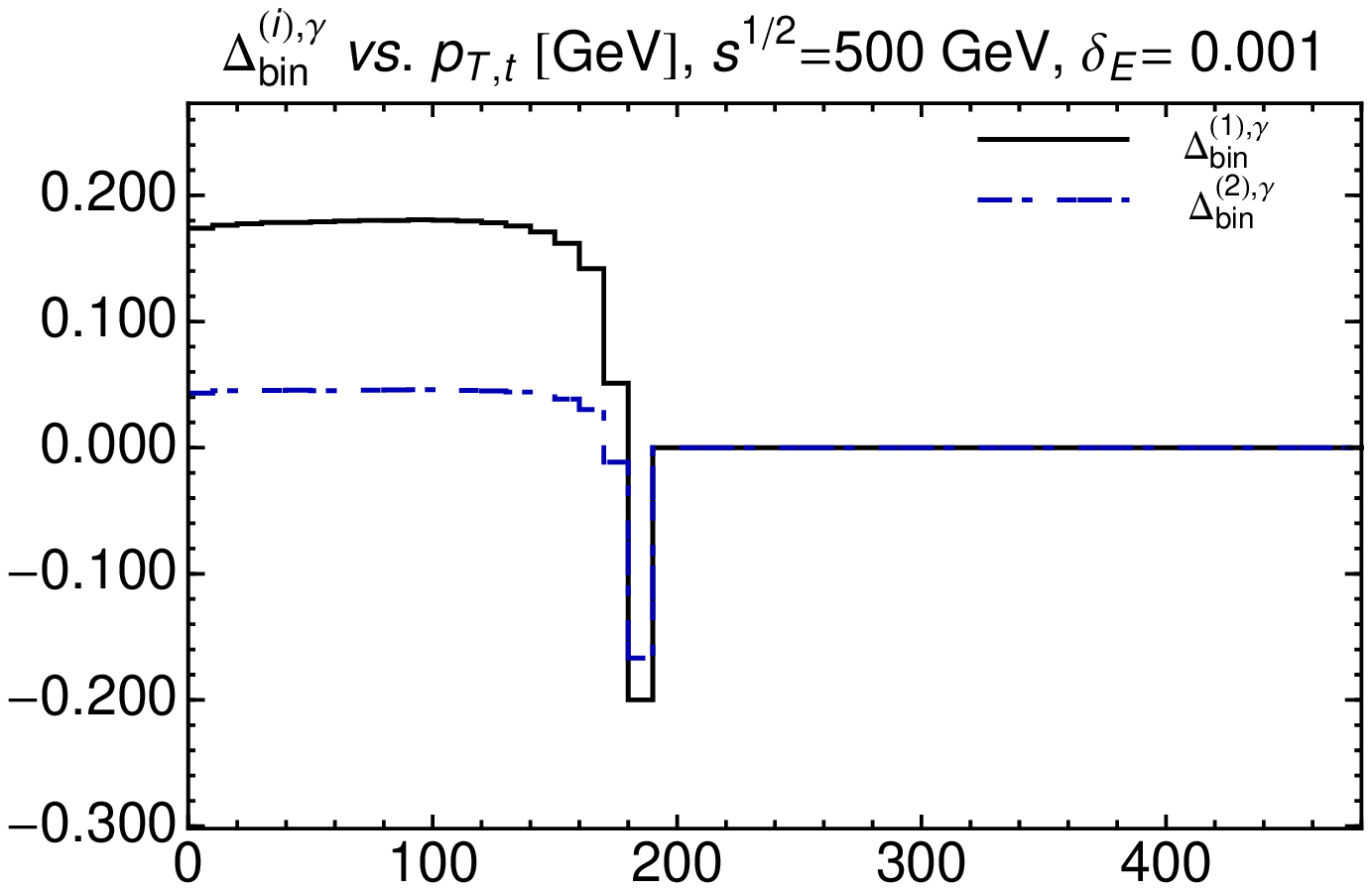}
  \includegraphics[width=0.4\textwidth]{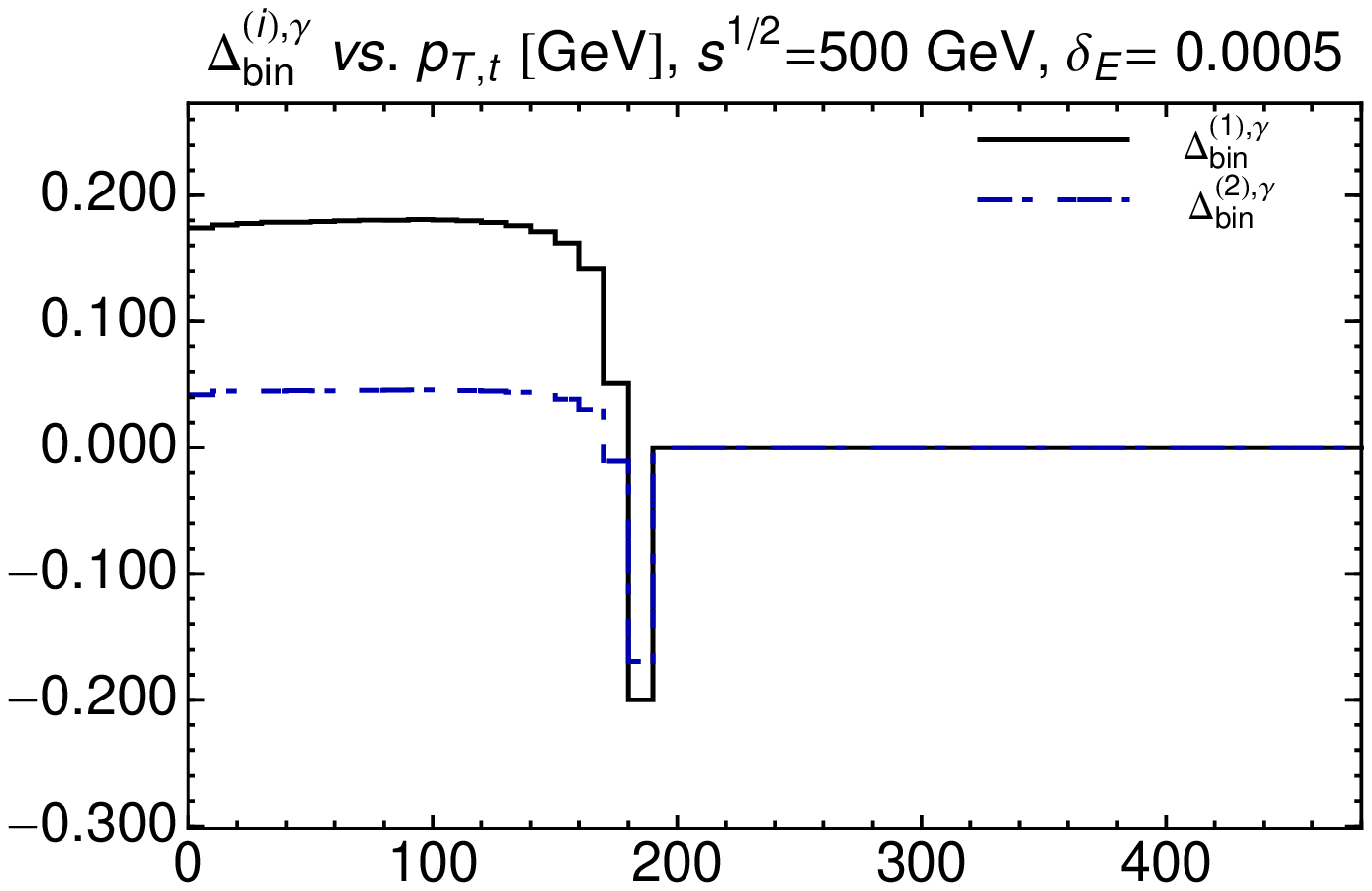}
\\
  \includegraphics[width=0.4\textwidth]{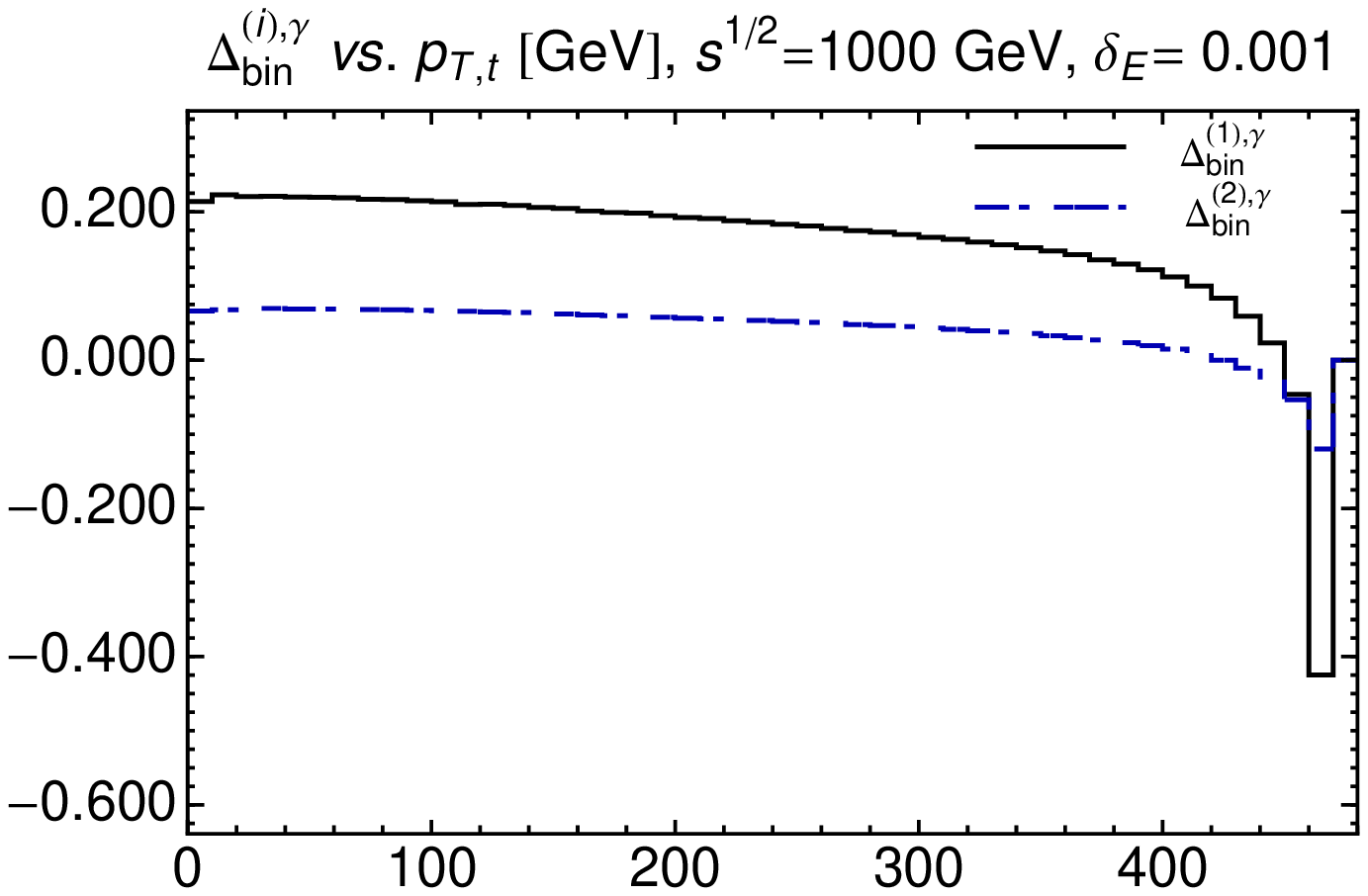}
  \includegraphics[width=0.4\textwidth]{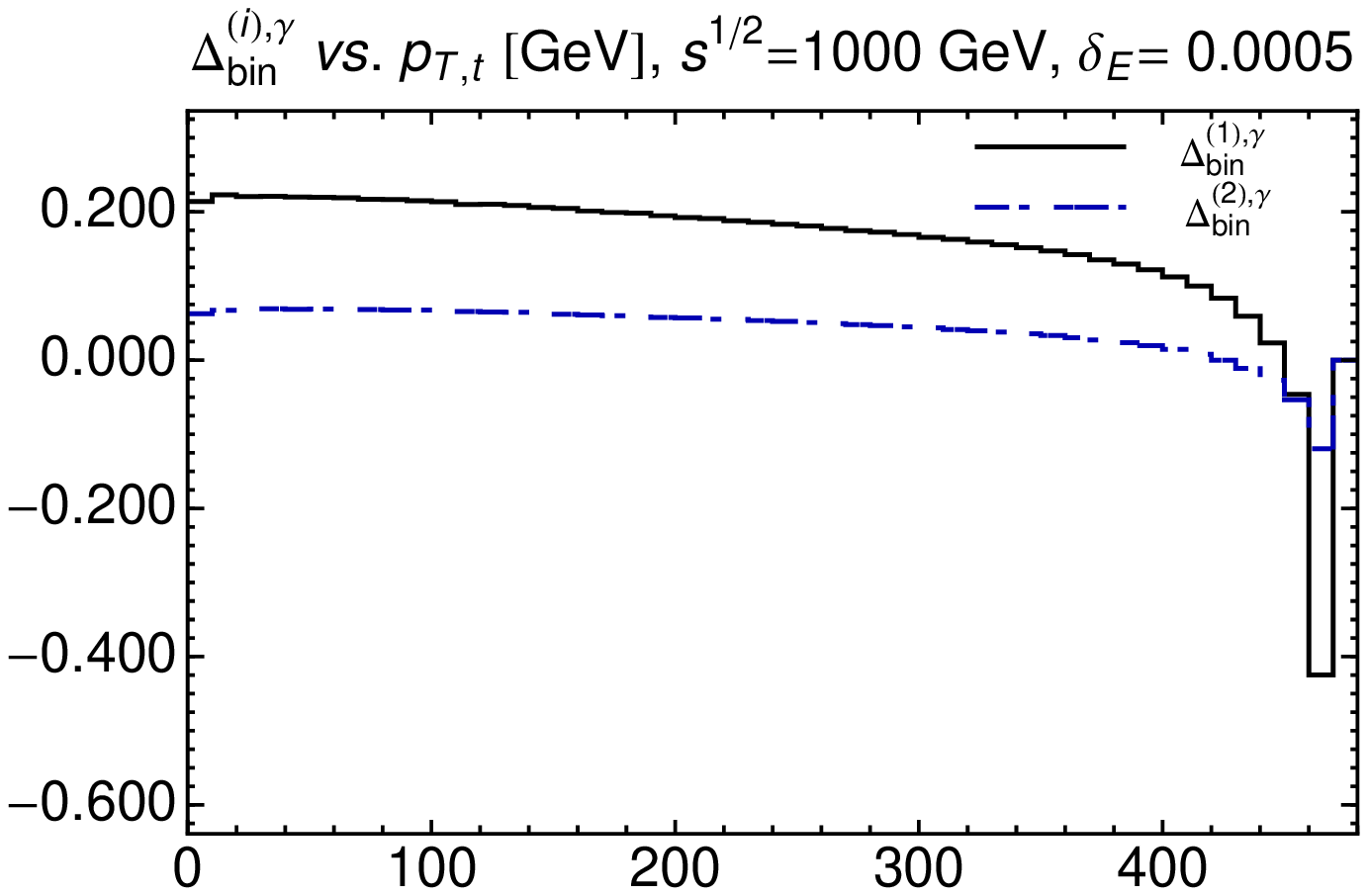}
  \end{center}
  \vspace{-1ex}
  \caption{\label{fig:dis2}
  NLO and NNLO corrections in different $p_{T,t}$ bins, $\Delta^{(1),\gamma}_{bin}$ and $\Delta^{(2),\gamma}_{bin}$,
  for different collision energies and different $\delta_E$ choices.}
\end{figure}

We can calculate fully differential distributions up to NNLO in QCD based
on the phase-space slicing method. At LO, there is only one non-trivial kinematic
variable, which we can choose either as cosine of the scattering angle between
the final-state top quark and the initial-state electron $\cos\theta_t$, 
or transverse momentum of the top quark with respect to the beam $p_{T,t}$.
Similar as the inclusive cross section, we can define the $\orda$ and $\ordb$ corrections
for each kinematic bin, $\Delta^{(1),\gamma}_{bin}$ and $\Delta^{(2),\gamma}_{bin}$, in analogy to Eq.~(\ref{eq:kfa}).
The results are shown in Fig.~\ref{fig:dis1} for $\cos\theta_t$ and~\ref{fig:dis2}
for $p_{T,t}$ distributions with collision energies of 350, 500,
and 1000 GeV. For each of them we plot the $\ordb$ corrections with two different
$\delta_E$ choices, $10^{-3}$ and $5\times 10^{-4}$. By comparing those two results
we can see very good stabilities of the $\ordb$ distributions for $\delta_E$ small enough $\sim$
a few $10^{-4}$, similar as the inclusive cross sections.

As can be seen from Fig.~\ref{fig:dis1}, both the $\orda$ and $\ordb$ corrections are flat for
$\sqrt s=350$ GeV where they are dominated by virtual corrections. The
$\cos\theta_t$ distribution is symmetric in forward and backward region for pure photon
contributions. For $\sqrt s=500$ GeV, the $\ordb$ corrections are slightly larger in region
of $|\cos\theta_t|\sim 1$ than central region, and are about 13\% of the $\orda$ corrections
in size. The $\ordb$ corrections for $\cos\theta_t$ distribution are totally negligible comparing
to the $\orda$ ones for $\sqrt s=1000$ GeV.

The transverse momentum distributions in Fig.~\ref{fig:dis2} show a different feature
comparing to the angular distribution since they are also affected by the energy spectrum
of the top quark. The real corrections pull the energy spectrum to the lower end and
thus the $p_{T,t}$ distribution as well. As shown in Fig.~\ref{fig:dis2}, both the $\orda$
and $\ordb$ corrections start as positive in low $p_T$ and then decrease to negative
values near the kinematic limits. The $\ordb$ corrections show a relatively larger
impact in the $p_{T,t}$ distribution.

\begin{figure}[h]
  \begin{center}
  \includegraphics[width=0.4\textwidth]{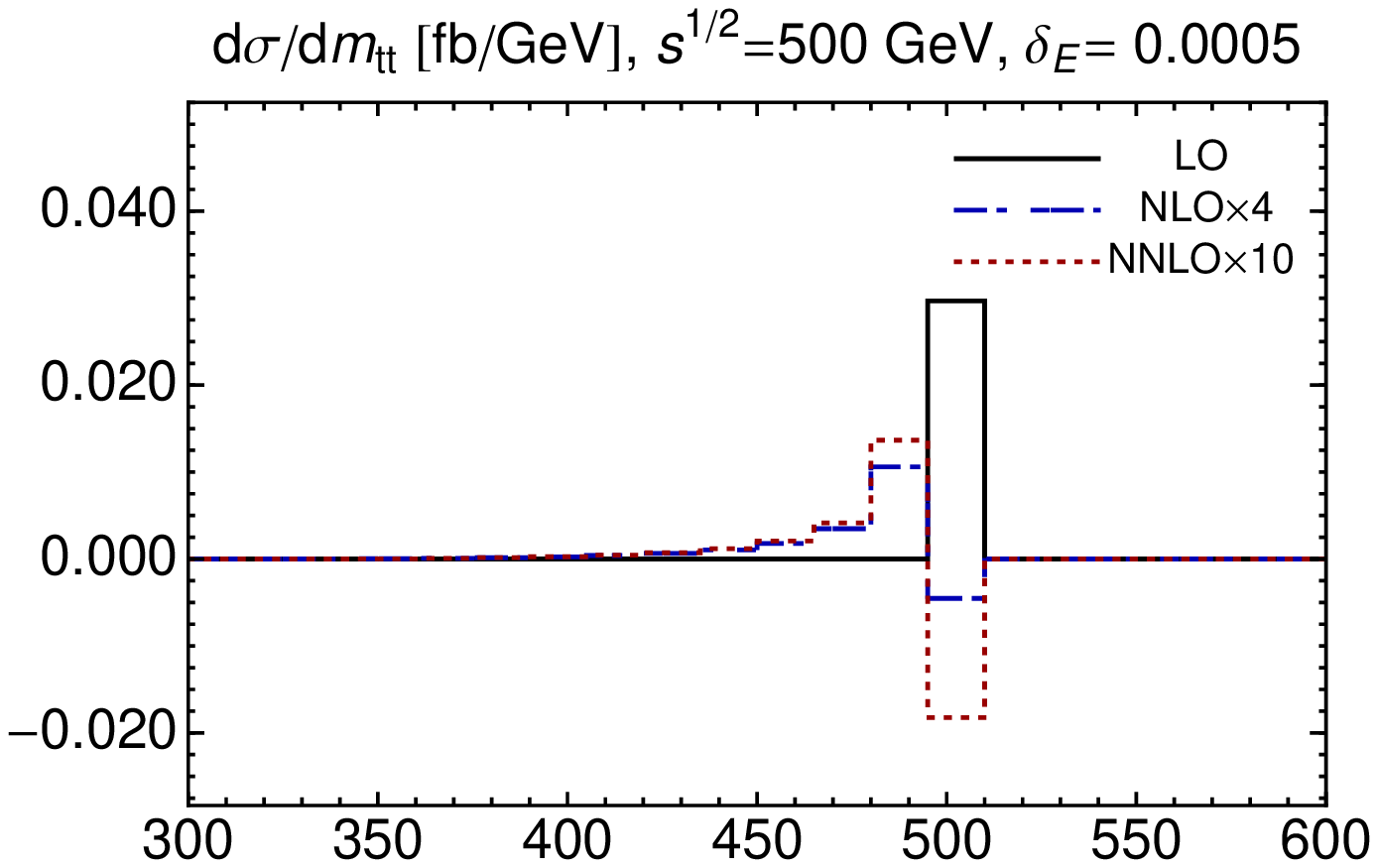}\hspace{0.5in}
  \includegraphics[width=0.4\textwidth]{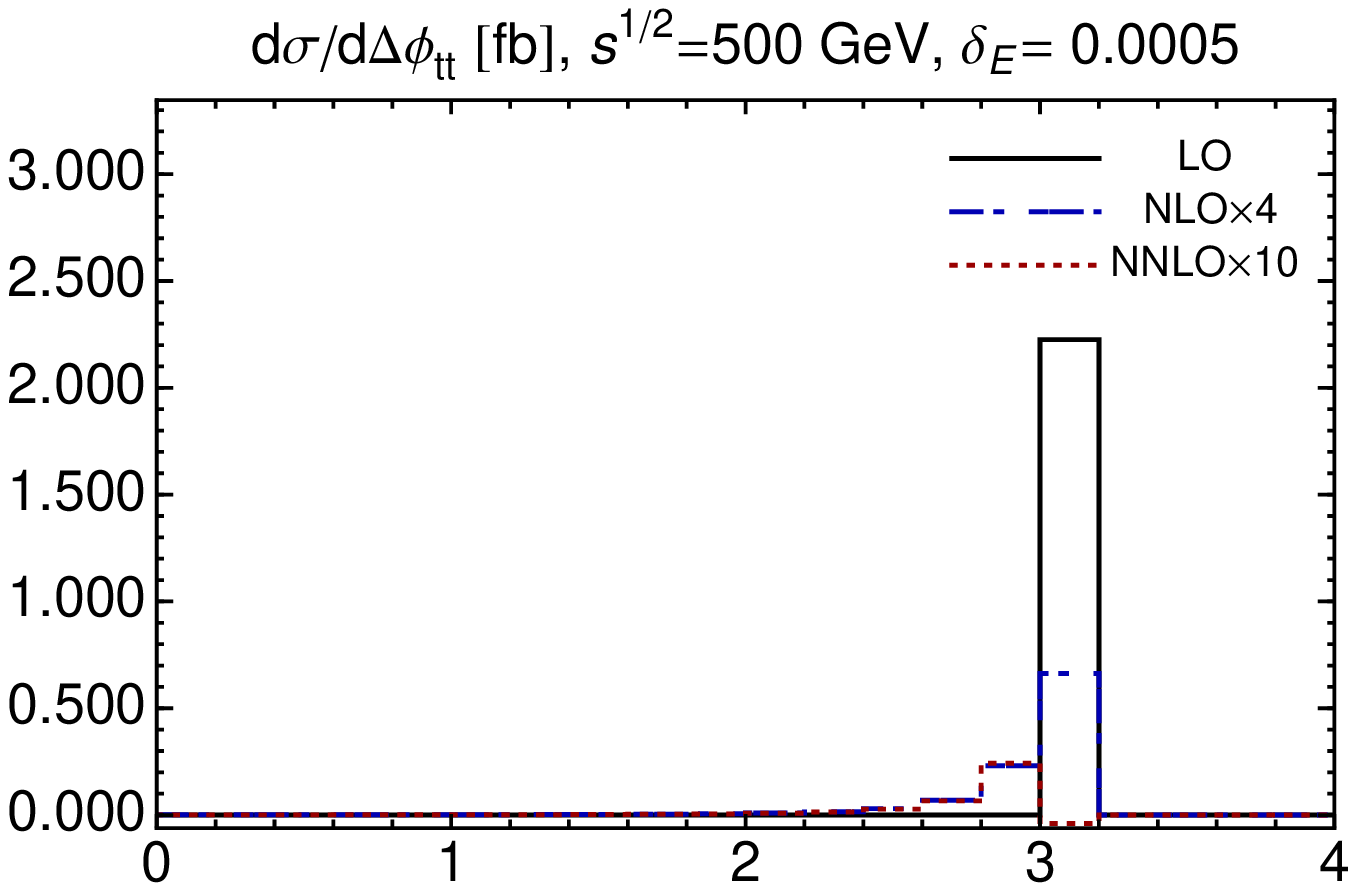}
  \end{center}
  \vspace{-1ex}
  \caption{\label{fig:dis3}
  Differential distribution, $d\sigma/dm_{t\bar t}$ on left, $d\sigma/d\phi_{t\bar t}$ on right,
  at the LO, $\mathcal{O}(\als)$ (multiplied by 4), and $\mathcal{O}(\als^2)$ (multiplied by 10).}
\end{figure}

Besides, we can also investigate distributions like, $\Delta\phi_{t\bar t}$, difference
of azimuthal angles of top and antitop quark, and their invariant mass, $m_{t\bar t}$.
Since they are both a delta function at the LO, our $\ordb$ corrections are effectively
NLO for those observables. We plot the LO distributions together with the $\orda$ and $\ordb$
corrections in Fig.~\ref{fig:dis3}. The corrections have been rescaled for comparison.
For bins with vanishing cross sections at the LO, we have compared our $\orda$ and $\ordb$
corrections with the calculations of $Q\bar Q+jet$ production up to NLO in~\cite{hep-ph/9705295} and
found very good agreement.

\section{C\lowercase{onclusion}}
\label{sec:conc}

To conclude, we have presented a fully differential NNLO QCD calculation for the photon exchange
contributions to electroweak top quark pairs production at $e^+ e^-$
colliders. Our calculations are based on a NNLO generalization
of the phase-space slicing method. Similar methods were introduced
some time ago to compute the $N_l$-dependent contributions to the total cross section~\cite{hep-ph/9505262,hep-ph/9707496}. To the
best of our knowledge, the results presented in this paper for the rest of
the color structures are new. Let us emphasize that we present various differential distributions as well at NNLO for the first time. 
Whenever possible, we have compared our results to existing analytical calculations. We find complete agreement with the known results, both in the
threshold~\cite{hep-ph/9712222,hep-ph/9712302,hep-ph/9801397,hep-ph/0001286}
and in the high-energy regimes~\cite{Gorishnii:1986pz,PHLTA.B248.359,hep-ph/9406299,hep-ph/9710413,hep-ph/9704222}. Although their calculation was beyond the scope of this work, the $Z$ exchange contributions
can be straightforwardly derived using the phase-space slicing technique discussed in this paper. The $Z$ exchange contributions are of fundamental phenomenological importance and will be treated in a future publication.

Inspired by the successful application of $q_T$ subtraction method of Catani and Grazzini~\cite{hep-ph/0703012}, recently there has been some interest and progress in applying the phase-space slicing method to NNLO QCD calculations.
For instance, top quark decay~\cite{1210.2808}, Drell-Yan production~\cite{1405.3607}, and Higgs production~\cite{1407.3773} have all been studied in schemes very similar to the one described in this work. 
This paper demonstrates that phase-space slicing can also be used to calculate top quark production processes, albeit at $e^+e^-$ colliders. 
Our calculation shows that fully differential NNLO corrections in $e^+ e^-$ annihilation are not much harder to obtain than typical NLO corrections to QCD processes once a good IR-safe observable has been defined
and the corresponding hard and soft functions are known.
In future work, it would be interesting to apply the phase-space slicing method to other NNLO QCD calculations relevant to the physics of future linear colliders and to generalize the method to allow for the treatment of
parton-initiated processes.

\begin{acknowledgments}
\noindent
We are grateful to A. Hoang and T. Tuebner for sharing their results
for the light quark contributions to the soft-virtual part of our calculation in the small mass
regularization scheme for comparison. We thank P. Nason and C. Oleari for providing us with the
numerical program used to perform the analysis discussed in ref.~\cite{hep-ph/9705295}. We are particularly indebted to R.~M.~Schabinger for numerious helpful discussions, and detailed feedback on the manuscript. We would also like to thank V. Hirschi for helpful discussions and to Y. Li for useful comments on the paper.
J.G. was supported by the U.S. DOE Early Career Research Award DE-SC0003870 and by the
Lightner-Sams Foundation. H.X.Z. was supported by the U.S. DOE under contract DE–AC02–76SF00515, and by the Munich Institute for
Astro- and Particle Physics (MIAPP) of the DFG cluster of excellence ``Origin and Structure of the Universe''.
\end{acknowledgments}

\end{document}